\documentclass{aa}
\begin{document}

       \thesaurus{06     
              (02.14.1;  
               08.01.1;  
               08.16.4;  
               08.03.1;  
               08.13.2;  
               08.02.1 )} 

 \title{ The heavy-element abundances of AGB stars and
  the angular momentum conservation model of wind accretion for barium stars }


\author{{Y. C. Liang$^{1,3}$ G. Zhao$^{1,3}$  B. Zhang$^{2,4}$ }
       }
       
 \offprints{Y. C. Liang, email: lyc@yac.bao.ac.cn}   

\institute{ $^1$Beijing Astronomical Observatory, 
               Chinese Academy of Sciences, Beijing 100012, P.R. China  \\
 $^2$Department of Physics, Hebei Normal University, Shijiazhuang 050016, P.R. China\\
 $^3$National Astronomical Observatories, Chinese Academy of Sciences, Beijing
 100012, P.R. China \\
 $^4$Chinese Academy of Sciences-Peking University Joint Beijing Astrophysical Center, 
 Beijing 100871, P.R. China 
      }
      
\date{Received  / Accepted}
\maketitle

\markboth{Y. C. Liang et al.: heavy-element abundances of AGB stars and
formation of Ba stars}{}

\begin{abstract} 
beginabstract: Command not found.

Adopting new s-process nucleosynthesis scenario and branch s-process path, 
we calculate the heavy-element abundances of solar metallicity 3$M_{\odot}$ 
thermal pulse AGB (hereafter TP-AGB) stars, 
and then discuss the correlation between heavy-element abundances and C/O ratio. 
$^{13}$C$(\alpha,n)^{16}$O reaction is the major neutron source, 
which is released in radiative condition during the interpulse period,
hence gives rise to an efficient s-processing that depends on the 
$^{13}$C profile in the $^{13}$C pocket.  
A second small neutron burst from the $^{22}$Ne source marginally 
operates during convective pulses over previously s-processed material diluted
with fresh Fe seed and H-burning ashes. 
The calculated heavy-element abundances and C/O ratio 
on the surfaces of AGB stars
are compared with 
the observations of MS, S and C (N-type) stars. The observations are 
characterized by a spread in neutron exposures: 0.5-2.5 times of the 
corresponding
exposures reached in the three zones of the $^{13}$C profile showed by Fig. 1
of Gallino et al. (1998). 
The evolutionary
sequence from M to S to C stars 
is explained naturally
by the calculated heavy-element abundances and C/O ratio. 

Then the heavy-element abundances on the surfaces of TP-AGB stars are used to
calculate the heavy-element overabundances of barium stars, which are generally
believed to belong to binary systems and their heavy-element overabundances
are produced by the accreting material from the companions (the former TP-AGB stars
and the present white dwarfs). To achieve this, firstly, the change 
equations of binary orbital elements
are obtained by taking the angular momentum conservation in place of the
tangential momentum conservation, and considering the change of $\delta r/r$
term; then the heavy-element overabundances of barium stars are calculated,
in a self-consistent manner, through wind accretion during successive pulsed
mass ejection, followed by mixing. 
The calculated relationships of heavy-element abundances to orbital periods $P$ 
of barium stars
can fit the observations within the error ranges. 
Moreover, the calculated
abundances of nuclei of different atomic charge Z, corresponding to 
different neutron exposures of TP-AGB stars, can fit 
the observational heavy-element abundances of 14 barium stars in the error ranges.
Our results suggest that the barium stars with longer orbital period 
$P>1600$ d may form through accreting part of the ejecta 
from the intrinsic AGB stars
through stellar wind, and the mass accretion rate is in the range
of 0.1-0.5 times of Bondi-Hoyle's accretion rate. 
Those with shorter orbital period 
$P<600$ d may be formed through dynamically stable late case C
mass transfer or common envelope ejection.

   \end{abstract}

%

\section{Introduction}
Since Burbidge et al. (1957) and Cameron (1957) published their creative 
studies, the nucleosynthesis theory has been developed in deep degree. In particular,
asymptotic giant branch (hereafter AGB) stars are very important 
to study element nucleosynthesis and the Galactic chemical evolution
because they 
synthesize significant parts
of slow process (hereafter s-process) neutron capture elements and $^{12}$C.
The products are taken out from the stellar interior,
He-intershell, to the surface by the third dredge-up (hereafter TDU) process, 
and then are ejected into interstellar medium with the progressive stellar 
wind mass loss. 

Our understanding of the AGB nucleosynthesis has undergone major revisions in 
these years. The earlier studies (Iben 1975; Truran \& Iben 1977)
illustrated that intermediate mass TP-AGB stars 
with $^{22}$Ne$(\alpha,n)^{25}$Mg neutron source 
(the typical neutron density is least of a order of $10^{9}-10^{10} n$ cm$^{-3}$)
are the suitable nucleosynthesis sites of s-process
elements. But new observations shed doubts on the above idea
(Busso et al. 1995 and references therein; Busso et al. 1999 and references therein).
Busso et al. (1995) and Lambert et al. (1995) 
demonstrated that the measured abundances of 
Rb/Sr, the products of the branch in the s-process path at $^{85}$Kr, imply a 
definitely lower neutron density (typical of the order of $10^{7}n$ cm$^{-3}$), 
which can be provided via the reaction $^{13}$C$(\alpha,n)^{16}$O at low temperature
in the He-intershell of low mass AGB stars.

Iben \& Renzini (1982a,b) indicated that a suitable mechanism operated in low
mass stars of low metallicity to allow the formation of a semiconvective layer,
hence the $^{13}$C pocket. The pocket is engulfed by the next convective pulse
where $^{13}$C nuclei easily capture $\alpha$ nuclei, release neutrons.
Hollowell \& Iben (1988) confirmed the possibility of formation of a consistent 
$^{13}$C pocket through a local time-dependent treatment of semiconvection.
However, the semiconvection mixing mechanism was not found to work 
for the $^{12}$C-enriched Population I red giants, 
like the peculiar stars of MS, S and C stars, showing
overabundances of s-process elements in their spectra. 

Straniero et al. (1995) investigated the effect of a 
possible mixing of protons into a thin zone at the top of the carbon-rich region 
during each dredge-up episode, hence the formation of $^{13}$C pocket.
They suggested that the $^{13}$C was completely burnt in the radiative condition, 
and the resulting s-process nucleosynthesis occurs during the quiescent 
interpulse period, instead of the convective thermal pulse. 
$^{22}$Ne$(\alpha,n)^{25}$Mg was still active for a very
short period during the convective pulse with minor influence on the 
whole nucleosynthesis.
Herwig et al. (1997) and Herwig et al. (1998) supported the formation of 
$^{13}$C pocket via hydrodynamical calculations.
 
Straniero et al. (1997) adopted the above new s-process nucleosynthesis scenario 
to calculate the s-process nucleosynthesis
of solar metallicity low mass AGB stars with $1 \leq M/M_{\odot} \leq 3$, 
and gave the detailed results.

Recently, Gallino et al. (1998) explained further and developed the aforesaid 
new scenario.
 They divided the $^{13}$C pocket, $q$ layer, into three zones 
 in the light of the distribution 
in the mass of hydrogen introduced in the $^{12}$C-rich intershell. The characteristic
neutron exposures in the three layers are different. 
Moreover, when the nucleosynthesis occurs in a radiative layer, only the nucleosynthesis
products are ingested into the convective thermal pulse,
which makes the classical concept of mean neutron exposure ($\tau_0$) 
become meaningless and the simple assumption of an
exponential distribution of the neutron exposure fail to account
for the complexity of the phenomenon (Arlandini et al. 1995, Gallino et al. 1998). 
Busso et al. (1999) reviewed this new s-process nucleosynthesis 
scenario in details. 
 
The spectral and luminosity studies of AGB
stars (including MS, S and N-type C stars) 
have shown that the M$\rightarrow$S$\rightarrow$C sequence is the result
that the low-mass AGB stars have undergone carbon synthesis, s-process
nucleosynthesis and the third dredge-ups (Lambert 1991). These stars are in the course
of experiencing thermal pulse stage, and the
original chemical abundances of their atmospheric envelope have been
modified by two mixing mechanisms, namely, the first dredge-up when they
became red giants and the third dredge-up when they became TP-AGB stars
(Boothroyd $\&$ Sackmann 1988a, b, c and d). 

The predicted evolutionary sequence
of M$\rightarrow$S$\rightarrow$C in the heavier$-$lighter
s-element abundances relationship (here and hereafter, 
'heavier' refers to the second metal peak 
elements: Ba, La, Ce, Nd and Sm etc.;  
'lighter' refers to the first metal peak elements: Y, Zr etc.)
and the heavy-element abundances$-$C/O relationship ($^{12}$C, together with
the s-process
elements, is dredged up from stellar interior during the third dredge-up)
are important to understand the nucleosynthesis and evolution of AGB stars. 
Because (1) they will
provide theoretical basis for the observed evolution of the sequence,
(2) they can check the available theories on the evolutions of AGB stars
(e.g., the beginning of the third dredge-up process, the mass and the chemical
abundance of the dredge-up material, the theory of s-process nucleosynthesis,
the stellar wind mass loss, and the formation
of carbon stars etc.). 

Busso et al. (1992) discussed the heavy-element abundances of
M, MS and S stars using the thermal pulse AGB model.
Busso et al. (1995) analyzed
the heavy-element overabundances of carbon stars under
the assumption that the dredge-up started after reaching the asymptotic
distribution (about the 20th pulse). 
It is difficult to calculate the AGB stars evolution and s-process
nucleosynthesis.
So there are few theoretical results to explain the  
M$\rightarrow$S$\rightarrow$C evolutionary
sequence based on the  
combination of the heavier$-$lighter s-element abundances ratio
and the C/O ratio,
though the observational abundances exhibit a certain regularity.
Zhang et al. (1998a) calculated the evolution of the
surface heavy element abundances and C/O ratio for a 3$M_{\odot}$ TP-AGB star with
initial metallicity 0.015, and gave interesting results.
But they adopted mean neutron exposure $\tau_0$ and unbranch s-process 
path, which have been revised in these years.

In the first part of this paper,    
we adopt the new s-process nucleosynthesis scenario (Straniero et al. 1995;
Straniero et al. 1997; Gallino et al. 1998; Busso et al. 1999 etc.),
and the branch s-process nucleosynthesis path to calculate the 
s-process nucleosynthesis of solar metallicity 3$M_{\odot}$ AGB stars.  
And then, we discuss the M$\rightarrow$S$\rightarrow$C
sequence based on the heavy-element abundances and C/O ratio. 

The importance of AGB stars nucleosynthesis is not only to explain 
the observational M$\rightarrow$S$\rightarrow$C sequence
but also to be responsible for the
origin of some other classes of stars with overabundances of
heavy-elements. 

Observations revealed that some stars with overabundances of
heavy-elements were not luminous to up to the stage of AGB. 
Following Lambert (1991),
the stars showing heavy-element overabundances are divided into two classes:
(1) intrinsic TP-AGB stars$-$they include MS, S and C (N-type) stars exhibiting
the unstable nucleus $~^{99}$Tc ($\tau_{1\over 2}=2\times 10^5 ~$yr) as
evidence that they are presently undergoing nucleosynthesis activity and
the third dredge-up, and (2) extrinsic AGB stars$-$they include the various
classes of G-, K-type barium stars and the cooler S, C stars where
$~^{99}$Tc is not observed.
It is generally believed that the extrinsic AGB stars 
belong to binary systems and 
their heavy-element overabundances come from
accretion of the matter ejected by the companions
(the former AGB stars, now evolved into white dwarfs) 
(McClure et al. 1980; Boffin \& Jorissen 1988; 
Jorissen et al. 1998; Jorissen \& Van Eck 2000; Jorissen 1999).
The mass exchange took
place about $1\times 10^6$ years ago, so the $^{99}$Tc produced in the
original TP-AGB stars have decayed. 
The accretion may either be 
disk accretion (Iben $\&$ Tutukov 1985) or common envelope ejection
(Paczynski 1976). Han et al. (1995) detailedly investigated 
the evolutionary channels for the formation of barium stars.
In this paper we will only discuss the stellar wind accretion scenario
because it is very important to explain the formation of barium stars 
(Boffin \& Jorissen 1988; Jorissen et al. 1998).

Boffin $\&$ Jorissen (1988) calculated
qualitatively the variation of orbital elements
caused by wind accretion in binary systems. 
They also estimated the heavy-element overabundances of barium stars. 
Subsequent
papers (Za$\check{c}$s 1994, Boffin $\&$ Za$\check{c}$s 1994) used similar
methods to calculate the overabundances, and interpreted the relationship 
between the heavy-element abundances and the orbital periods
of barium stars.

Some important conclusions have been drawn in the theory of wind accretion,
but the 
previous calculations on orbital elements were not very reliable 
because of the neglect of the $\delta r/r$ term 
($r$ represents the distance between the two components of the binary system
), and using the 
tangential momentum conservation  
(Chang et al. 1997 and references therein). 
For the rotating binary system with wind mass loss, 
the total angular momentum conservation is more reasonable
than tangential momentum conservation.   
Also, the previous calculations of heavy-element overabundances used 
the 'step-process'
(Boffin $\&$ Jorissen 1988, Boffin $\&$ Za$\check{c}$s 1994), 
which means that 
the overabundance factor changes at one single instant from 1 to $f$
($f$ is the relative ratio of the heavy-elements to the solar abundances, 
and differs for different elements. 
Earlier calculations used the mean $f$ value of carbon stars), 
and then keeps the value until the end of the AGB phase. 
However, after the start of the TDU,
the overabundance factor $f$ of the intrinsic AGB stars changes during 
successive dredge-ups. It is after a number of dredge-ups 
that the C/O ratio in the
outer envelope of the intrinsic AGB star reaches the value 1, 
which means that the AGB star becomes a carbon
star. The heavy-element overabundances of the barium star should be 
caused by the successive pulsed accretions and mixing. 

According to the analysis to
the orbital elements of barium and S stars,
Jorissen \& Mayor (1992) presented the evolutionary pathways of binaries
 leading to barium and S systems. They concluded that the binary
systems with longer orbital period formed through wind accretion and 
those with shorter orbital period formed via Roche lobe overflow. 
But the specific range of orbital
period was not presented. Jorissen et al. (1998) analyzed the orbital
elements of a large sample of binary systems to 
give insight into the formations of barium and Tc-poor S stars.
They suggested that barium stars with orbital period $P$$>$1500 days
formed through wind accretion scenario. Zhang et al. (1999) suggested
that the barium stars with $P$$>$1600 days formed 
via wind accretion according to their model. 
Liu et al. (2000) discuss the evolutions of orbital elements
of barium stars from normal G, K giants.

Besides the orbital elements, the 
heavy-element abundances of barium stars have been discussed 
in some literatures.
Busso et al. (1995) discussed
the observational heavy-element abundances of barium stars.
And a more detailed analysis of the abundance distributions 
for five stars has been performed 
using the method of mixing the accretion mass with
the envelope mass. 
But the effect of
mass accretion and the changes of orbital elements were not considered.
Chang et al. (1997) and Liang et al. (1999) attempted to explore 
the relationships between the heavy-element overabundances 
and orbital elements
of bariums stars using the binary accretion scenario,
but with the shortcomings: or using the tangential 
momentum conservation, or adopting the old nucleosynthesis scenario 
of TP-AGB stars. 

In the second part of this paper, we firstly 
reduced the variation equations of orbital elements based on
the 
angular momentum conservation model of wind accretion,
then we calculate the heavy-element overabundances of barium stars
via successive pulsed accreting matter enriched heavy-elements
from the intrinsic AGB stars, and mixing the matter with
their envelopes. 

This paper is organized as follows.
The observational data of MS, S, C (N-type) stars 
and barium stars are given in Sect. 2. 
In Sect. 3, we present the model and the main parameters of AGB stars 
nucleosynthesis and 
the angular momentum conservation model of wind accretion scenario
for barium stars. 
Sect. 4 illustrates and analyzes our results. 
We conclude and discuss in Sect. 5.

\section{Observations}
The abundance data of the heavy-elements, 
C and O elements of MS and S
stars are taken from Smith $\&$ Lambert (1985, 1986, 1990).
The listed values of [N/Fe] (=log(N/Fe)$_{\rm star}-$log(N/Fe)$_{\rm sun}$) have
typical errors 0.1-0.2 dex, and N represents a kind of
s-process element.
Indeed, for the observations in the series papers of Smith \& Lambert, 
the accuracy of the iron peak elements (Ti, Fe and Ni) abundances and the
heavy-elements Y, Zr, Nd abundances were better. But the abundance of Nd element
is only at 50$\%$ of s-origin, it is not reasonable to regard Nd as the
representative of heavier s-elements (Busso et al. 1992; Zhang et al. 1998a),
and the reasonable representative should be the mean abundance of 
Ba, La, Nd and Sm (Busso et al. 1995).
In order to do this we suggest that the initial abundance of Nd have to be
increased by 0.11 dex before the data of MS and S giants are compared with
carbon stars (Busso et al. 1995). Y, Zr are the representatives of lighter
s-elements. For C and O elements, the accuracy is higher at 0.1 dex in 
Smith $\&$ Lambert (1990).
Considering that, at present, the sample with both observed C/O ratios and
heavy-element abundances is rather limited and there is a close relationship 
between the Tc-no MS, S stars (stars that contain no Tc) and the thermal 
pulse AGB stars (Busso et al. 1992; Jorissen $\&$ Mayor 1992), 
we can usefully include the Tc-no stars in our study of evolutionary 
tendency of the M$\rightarrow$S$\rightarrow$C sequence
(Busso et al. 1992).

The abundance data of heavy-elements 
of the 12 carbon stars (N-type) are taken from Utsumi (1985), 
with typical errors of 0.4 dex in [N/Ti], which are dealt 
like in Busso et al. (1995).
The corresponding abundances of C, O elements are taken from 
Lambert et al. (1986).
 
The heavy-element abundances of 12 barium stars are 
taken from Za$\check{c}$s (1994). 
For $\zeta$ Cap and HR 774, the heavy-element abundances
are from the same sources as in Busso et al. (1995). 
The sun, instead of $\epsilon$ Vir, 
is chosen to be a comparison standard,
namely, [N/Fe]=log(N/Fe)$_{\rm star}-$log(N/Fe)$_{\rm sun}$. 
The abundances of $\epsilon$ Vir to the sun are
taken from Tomkin $\&$ Lambert (1986). The new solar system abundances 
are taken from Anders \& Grevesse (1989).
The overabundances of s-process elements in
barium stars are characterized by a mean value of
[s/Fe]=([Y/Fe]+[Ba/Fe]+[La/Fe]+[Ce/Fe]+[Nd/Fe])/5.0.
The typical error is 0.2 dex (Boffin $\&$ Za$\check{c}$s 1994). 
Zr has been excluded
from this mean value, since the abundance of Zr is poorly related to 
other heavy-elements in the sample
(Boffin $\&$ Za$\check{c}$s 1994).
The corresponding observational data of orbital elements, 
orbital period $P$ and eccentricity $e$, are taken from 
McClure $\&$ Woodsworth (1990),
Boffin $\&$ Za$\check{c}$s (1994) and Jorissen et al. (1998).
 
\section{The model and the main parameters}

\subsection * {3.1 The nucleosynthesis and mixing in AGB stars}
AGB star consists of three layers: a C-O core of degenerate
electron, a He and H double burning shell, and a convective envelope.
Heavy-elements synthesized in the He-intershell are dredged up to the convective envelope
by the convection that appears in the cooling contraction after a thermal pulse,
and show up as the observed overabundances of the heavy-elements and $^{12}$C.
With the evolution of the star, 
its core mass steadily increases with increasing thermal pulses, while
the mass of the convective envelope decreases correspondingly via
wind mass loss, until it is completely exhausted, while marks the life span of
the thermal pulse AGB star.

During every thermal pulse, we calculate the s-process nucleosynthesis.
$^{56}$Fe is taken to be the seed nucleus for the s-process. The reaction
$^{209}$Bi(n,$\gamma$)$^{210}$Bi$(\alpha)$$^{206}$Pb($^{210}$Bi has an
$\alpha$-decay half-life of 5.01 days) terminates the reaction chain of the slow
process of neutron capture. The branch s-process path 
is adopted in the calculation. 
The neutron capture cross sections of most of nuclei,
are taken from Beer et al. (1992).
The $\beta$-decay rates and electron capture rates of nuclei are
taken from Takahashi \& Yokoi (1987). More recent values
of some nuclei are taken from the same literatures as 
those cited by Gallino et al. (1998).
For the initial abundances, we use standard red giant abundances, which differ
from the solar abundances only by a constant factor:
$N_i=N_{i\odot}\times Z/Z_{\odot}$. At present, the solar system abundances are 
the most detailed abundance distribution obtained by us. 
According to the results of Straniero et al. (1997) and Gallino et al. (1998), 
the main parameters adopted in calculation are as following.

(1) Core mass $~~$ At the onset of the thermal pulse, the mass
of the C-O core is 0.572$M_{\odot}$. With successive pulses, 
both the H burning
and He burning shells move outwards (in mass coordinates)
and so does the C-O
core. The core mass is up to  0.611$M_{\odot}$ at the 8th pulse, 
during which the
third dredge-up begins. The variation of the core
mass influence directly the wind mass loss. The core masses of stars
in different
TP periods are taken from table 4 of Straniero et al. (1997).

(2) Mass loss through wind $~~$ 
Wind mass loss is one of the most important ingredients 
in the computation of both the
nucleosynthesis occurring in TP-AGB stars and the wind accretion 
on the secondary
component of a binary system. Here, we adopt the mass loss rates given
by Straniero et al. (1997) using Reimers formula (Reimers 1975) 
based on the TP-AGB stellar models
(their table 4).

The mass lost during the time $\Delta t$, the interpulse period, is
\begin{equation}
\Delta M = -\dot M\Delta t,
\end{equation}
which is taken from table 4 of Straniero et al. (1997).

(3) Dilution Factor $~~$ 
The physical process that influences
significantly the surface abundance is the third dredge-up. The dilution factor
$f$ is defined as the ratio of the mass dredged in every pulse to the mass
of the envelope after the third dredge-up begins, namely,
\begin{equation}
f=\frac{\Delta M_{\rm TDU}}{M_{\rm env}+\Delta M_{\rm TDU}} ,
\end{equation}
where $\Delta M_{\rm TDU}$ is the mass dredged up from He-intershell 
at every pulse since TDU begins, and
$M_{\rm env}$ is the mass of the envelope. 
The cumulative effect of the various
TDU, $\sum \Delta M_{\rm TDU}$, have been given in Fig. 9a of Gallino et al. (1998).

(4) Overlap factor $~~$ 
The overlap factor $rr$, $rr=1- \Delta M_{\rm c}/M_{\rm CSH}$,
is the ratio of
the number of a kind of nuclide that is exposed to neutron  
in two successive
pulses to the number exposed in the previous pulse, that is, 
the overlap fraction of
the convective shells between the two successive pulses. 
Where $\Delta M_{\rm c}$ is the increment of the core mass during the previous 
interpulse period, 
and $M_{\rm CSH}$ is the mass of the convective inter-shell.
The variation of $rr$ with
core mass was given by Iben (1977) and Renzini $\&$ Voli (1981).
According to suggestion of Straniero et al. (1995), 
$rr$ steadily decreases with increasing core mass. 
In this paper, 
we adopt the values given by Fig. 9c of Gallino et al. (1998).

(5) Neutron source $~~$ 
$^{13}$C$(\alpha,n)^{16}$O reaction is the major neutron source
of s-process nucleosynthesis, which is released
in radiative conditions at $T_8 \sim 0.9$ during the interpulse period, hence 
gives rise to an efficient s-processing that depends on the $^{13}$C profile 
in the $^{13}$C pocket, $q$ layer. 
The layer is divided into
three zones, in which the neutron exposures change 
with the various interpulse
periods (Gallino et al. 1998, see their Fig. 6). 
Then the products of nucleosynthesis are ingested into the convective 
thermal pulse, 
and mixed with s-process material already from the previous pulses, 
and with the H-burning ashes from the below H shell. 
A second small neutron burst 
from the $^{22}$Ne source operates during convective pulses
due to the high neutron density and high temperature. The $^{22}$Ne source to 
the total
neutron irradiation is small, ranging from $\sim$0.002 mb$^{-1}$ at the 9th pulse to 
$\sim$0.05 mb$^{-1}$ in the latest pulses, where the average effective temperature 
is $\sim$23 keV.  

(6) Light neutron poisons $~~$
Some isotopes can consume neutron 
produced by the neutron source isotopes, $^{13}$C or $^{22}$Ne,
through some reactions. They are named as neutron poisons.
For example, the isotope $^{14}$N through its resonant
channel $^{14}$N$(n,p)^{14}$C, 
$^{26}$Al in its long-lived ground state synthesized by the reaction 
$^{25}$Mg$(p,\gamma)^{26}$Al$(p,\gamma)^{27}$Si (Gallino et al. 1998 
and references therein). However, in the previous convective pulses, 
$^{14}$N is destroyed in the reaction chain leading to $^{22}$Ne production, 
and a large fraction of $^{26}$Al has undergone substantial depletion
by neutron capture. The above discussion makes it clear that compared to 
a convective burning scenario, 
the radiative s-process is much less affected by the filtering
effect of light neutron poisons (Gallino et al. 1998). 
So in the radiative s-process 
nucleosynthesis calculation, we neglect the influence of these 
light neutron poisons 
(the slightly effects on nucleosynthesis calculation 
caused by this will be further discussed in section 4.1).

(7) Value of C/O $~~$
At the beginning of the AGB phase, the C/O on the surface 
of a low mass stars
is approximately 0.31, and lower than the initial (solar) value (approximately 0.45)
because it has been modified by the first dredge-up process, 
which reduces the $^{12}$C
abundance of the surface by about 30\%.
With the third dredge-up process,
$^{12}$C of He-intershell is mixed into the stellar surface
with the mixing of s-process material. So the C/O ratio on the surface 
increases gradually.  
In calculation, we adopt the C/O ratio from table 4
of Straniero et al. (1997). 

\subsection * {3.2 The heavy-element overabundances of barium stars}

\subsubsection * {3.2.1 The angular momentum conservation model of wind accretion }

For the binary system, the two components
(an intrinsic AGB star, the present white dwarf, with mass $M_{1}$,
 and a main sequence star,
the present barium star, with mass $M_{2}$) rotating around
the mass core C, so the total angular momentum is conservative in the mass
core reference frame. If the two components exchange material through wind
accretion, the angular momentum conservation of total system is showed by:
\begin{equation}
\Delta(\mu r^{2}\dot{\theta})=\omega r_{1}^{2}(\Delta M_{1}+\Delta M_{2})
+r_{2}v(\Delta M_{1}+\Delta M_{2}),
\end{equation}
where $\mu$ is reduced mass,
 and $r$ is the distance from $M_{2}$ to $M_{1}$.
 $r_{1}$, $r_{2}$ are the distances from $M_{1}$, $M_{2}$ to the
mass core C respectively.
$\omega$ ($={2\pi}/P$, where $P$=2$\pi A^2(1-e^2)^{1 \over 2}/h$
is orbital period) is angular velocity.
$v$ is an additional effective velocity defined through the
angular momentum variation in the direction of orbital motion of component 2.
The first term on the right side of the equal-sign is the angular momentum
lost by the escaping material, and the second term is the additional angular
momentum lost by the escaping material. 

Using the similar method to that
adopted by Huang (1956), Boffin $\&$ Jorissen (1988) and Theuns et al. (1996), 
considering the angular momentum conservation of total system and
not neglecting the square and higher power terms of eccentricity,
we can obtain the change equations of the orbital elements:

\begin{eqnarray}
 \frac{\Delta A}{A} & = &  - 2(1-e^2)^{1\over2} \left [ \frac{\Delta M_1}{M_1}+\frac{\Delta M_2}{M_2}
                 -  \frac {\Delta M_1+ \Delta M_2}{M_2} \frac {v} {v_{\rm orb}} \right ] \nonumber \\
                 & + &  2(1-e^2)^{1\over2} \frac {M_2 (\Delta M_1+ \Delta M_2)}{M_1(M_1+M_2)} \nonumber \\
                 & + & [2(1-e^2)^{1\over2}-1] \frac {\Delta M_1+ \Delta M_2}{M_1+M_2},            
\end{eqnarray}

\begin{eqnarray}
  \frac{e\Delta e}{1-e^2} & = &  [1-(1-e^2)^ {1\over2}]
                 \left [ \frac {\Delta M_1}{M_1}+\frac {\Delta M_2}{M_2}         
                 - \frac {\Delta M_1+ \Delta M_2}{M_1+M_2} \right ] \nonumber \\
               & - & [1-(1-e^2)^ {1\over2}]\frac {M_2 (\Delta M_1+ \Delta M_2)}{M_1(M_1+M_2)}  \nonumber \\
               & - & \frac {3e^2}{2(1-e^2)^{1\over2}} 
                     \frac {\Delta M_1+ \Delta M_2}{M_2} \frac {v}{v_{\rm orb}} ,
\end{eqnarray}

where $A$ is the semi-major axis of the relative orbit of component 2 around 1, 
and $e$ is the eccentricity (more details can be found in Appendix and Liu et al.
(2000)).
Here, we take $v=0$ (Boffin $\&$ Jorissen 1988). 

For the mass accreted by the barium star,
we use the Bondi-Hoyle (hereafter B-H) accretion rate (Bondi \& Hoyle 1944; 
Theuns et al. 1996; Jorissen et al. 1998):
\begin{equation}
\Delta M_{2}^{\rm acc}=- \frac {\alpha}{A^{2}} \left (\frac{GM_{2}}{v_{\rm ej}^{2}} \right )^{2}
 \left [\frac{1}{1+(v_{\rm orb}/v_{\rm ej})^{2}} \right ]^{3\over2} \Delta M_{1} ,
\end{equation}
where $\alpha$ is a constant expressing the accretion efficiency, and its value is
taken from Theuns et al. (1996). Theuns et al. (1996) indicated
that taking $\alpha$ between 0.5 and 1, as suggested by the numerical 
simulations of Ruffert \& Arnett (1994), 
the actual accretion rate deduced from the smoothed particle hydrodynamics 
(hereafter SPH) simulation in the $\gamma=1.5$ case was thus about 
10 times smaller than that predicted by the B-H formula. 
Here we take $\alpha$=1 for the B-H formula.
In fact, Boffin \& Za$\check{c}$s (1994) suggested that 
the actual accretion rate is between
0.1 and 1 times of B-H rate. We  
take 0.15 times of the B-H rate as the actual accretion 
rate for our standard case. 
$v_{\rm ej}$ is the wind velocity, and $v_{\rm orb}$
is the orbital mean velocity. 
After fixing the initial conditions, for the mass
$\Delta M_1$, ejected at each pulse by the primary star, 
we can solve the Eqs.(4)-(6)
for $\Delta M_2$, the accreted mass by the secondary star. 
And then, the heavy-element abundances on the surface of
barium stars can be calculated. 

\subsubsection * {3.2.2 The heavy-element overabundances of barium stars}

The calculation is completed by two separated steps. Firstly, adopting
the theory of s-process nucleosynthesis and the latest TP-AGB model, 
we calculate the degree
of the overabundances of the intrinsic AGB star (the present white dwarf) at
each ejection. Then, combining the accreting matter predicted by the model of
wind accretion on successive occasions and mixing, 
we calculate the heavy-element
overabundances of the barium star. The overabundance factor for nuclide
$i$ on the secondary star, $g_i$, is given by
\begin{equation}
g_i=\frac{M_{\rm env}g_{i}^{0}+\sum\limits_{n=1,m} f_{i}^{n} M_{\rm acc}^{n}}
{M_{\rm env}+\sum\limits_{n=1,m}M_{\rm acc}^{n}} ,
\end{equation}
where $M_{\rm env}$ is the mass of the outer envelope of barium star,
$M_{\rm acc}^{n}$ is the mass accreted by barium star during the period of
the $n$-th ejection of the intrinsic star, and $f^{n}_{i}$ is the 
overabundance factor
of the nuclide $i$ of the intrinsic AGB star on that occasion. $m$ is the total
ejecting number undergone by the intrinsic AGB star. $g_{i}^{0}$ is
the overabundance factor of nuclide $i$ of barium star before the mass
accretion. In the above formula we have assumed that the accreting matter has
already completely mixed with the outer convective envelope of barium
star.

We take as standard case: $M_{1,0}$=3.0$M_{\odot}$,
$M_{2,0}$=1.3$M_{\odot}$, $v_{\rm ej}$=15${\rm ~km~s^{-1}}$ 
and 0.15 times of the Bondi-Hoyle's accretion rate.

\section{Results and analysis}

\subsection * {4.1 The intrinsic AGB stars}

Using the recent evolutionary model and nucleosynthesis scenario of intrinsic
TP-AGB stars (Straniero et al. 1995; Straniero et al. 1997; Gallino et al. 1998),
we calculate the heavy-element abundances on the surface of solar metallicity
3$M_{\odot}$ AGB star with
wind mass loss. The TDU begins at the 8th pulse.
The average heavy element abundances on the surface
are given in Fig. 1. 

\begin{figure}
\input epsf
\epsfverbosetrue
\epsfxsize 8.8cm 
\epsfbox{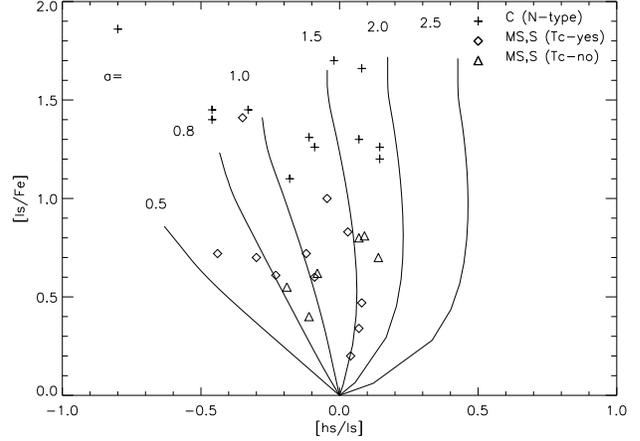} 
\caption{~Comparison of the theoretical predictions of surface 
heavy-element abundances for different neutron exposures (solid lines) with 
observations of intrinsic AGB stars. The numbers of '$a$' represent the times 
of the corresponding exposures in the $^{13}$C profile suggested by the 
Fig. 1 of Gallino et al. (1998). }
\end{figure}

In Fig. 1, the abscissa represents the average logarithmic ratio of 
the abundances of the heavier s-elements Ba, La, Nd, and Sm ('hs')
to the lighter s-elements Y and Zr ('ls'),
which depends on the neutron exposure.
The ordinate is the logarithmic enhancement in 'ls' with respect to iron
in the envelope, 
which is closely related
to the synthesis in the He-intershell and the dilution factor. 
The different curves represent the results of different neutron exposures.

Indeed, the neutron exposures of $^{22}$Ne in different thermal pulses
have been showed in Sect. 3.1. 
In calculation, we adjust the
neutron exposure of $^{13}$C source. Gallino et al (1998) suggested the
$^{13}$C profile in $^{13}$C pocket by their Fig. 1,  
namely their 'standard' case. Then they showed that the resulting 
heavy-elements
abundances were nonsolar by their
Fig. 14,
while simply increasing the previously adopted abundance of $^{13}$C
by a factor of 2 could reproduce the main component of solar system 
(see their Fig. 16).
Here, 
instead of the abundance of $^{13}$C,
we adjust the total neutron exposure caused by $^{13}$C source
to calculate the nucleosynthesis of AGB stars. 
Thus, the different values of '$a$' in our Fig. 1 represent the different 
times of the actual total neutron exposures to the 'standard' case  
suggested by Gallino et al. (1998). 
Our result of $a$=1.0 corresponds to the 'standard' case of
Gallino et al. (1998), 
and the $a$=1.5 case corresponds to their results of
increasing the previously adopted abundance of $^{13}$C by a factor of 2.
In reality, Gallino et al. (1998) have illustrated this times relationship (1.5 vs. 2) 
in the description and discussion about their 
Figs.14, 15 and 16.

Our Fig. 1 shows that 
our results are very similar to those of Gallino et al. (1998).
The curve with $a$=1.0 exhibits clearly that 
the abundances of the heavier
nuclei are lower than those of the lighter nuclei in the 23rd interpulse phase 
like Gallino et al. (1998),
namely [hs/ls]$<$0.0 (the abscissa), which means that
this case can not reproduce the solar abundance distribution 
due to a lower production for heavier nuclei. 
While the curve with $a$=1.5 produces
[hs/ls]$\sim$0.0 in the 23rd interpulse phase, 
which means that the abundances of the heavier and the lighter nuclei 
are solar like. 
So, our 'average' heavy-element abundances on the surfaces of AGB stars 
can fit the detailed abundance distribution of nuclei obtained by Gallino et
al. (1998). 
In addition, we should note that, because we neglect the effects of light 
neutron poisons on nucleosynthesis
in calculation, our '$a$' values are slightly higher than the
corresponding case of Gallino et al. (1998). 

We calculate some other curves with different $a$ values, which are 
given in Fig. 1 too. The results show that all of our available observational
data are compatible with the range 0.5$\leq$a$\leq$2.0.
The evolutionary
curves for three cases ($a$=1.0, 1.5, 2.0) move
upward as the dredge-ups proceed, and from the region of MS, S stars to
reach the region of the carbon stars
(symbol plus).
And the larger neutron exposure, 
the higher [hs/ls] ratio will be, which means
the large neutron exposure benefits the production of the heavier
s-process elements. 

$^{12}$C isotope in the He-intershell of TP-AGB stars,
together with the s-process elements,
is dredged-up
and mixed to the envelope by the TDU process, 
which causes that the C/O ratio on the surface increases 
gradually with the thermal pulses. The increase correlates to the
overabundances of heavy-elements.
Our Fig. 2 displays the relationship of heavy-element abundances to C/O 
ratio. The values of '$a$' are relevant to the values in Fig. 1. 
From the beginning of TDU, the values of C/O ratio and heavy-element 
abundances increase gradually. After some dredge-ups, C/O becomes greater
than 1, that is, the stars become carbon stars ($a$=1.2, 1.5, 2.5).
The different curves show that the higher
neutron exposure, the more heavy elements are
produced. 

\begin{figure}
\input epsf
\epsfverbosetrue
\epsfxsize 8.8cm  
\epsfbox{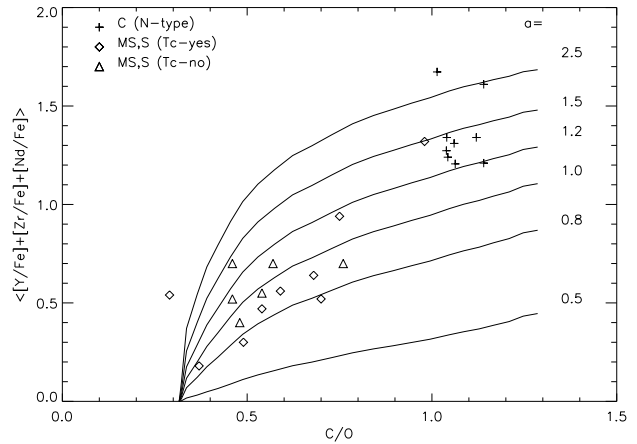} 
\caption{~Comparison of theoretical predictions of the relationships between
heavy-element abundances and C/O ratios for different neutron exposures
(solid lines) with observations of intrinsic AGB stars.
The numbers of '$a$'represent the same meanings as in Fig. 1.}
\end{figure}

The results of Figs.1 and 2 illustrate that, adopting the new TP-AGB nucleosynthesis
scenario, choosing the reasonable parameters, 
we can explain the observed heavy-element abundances and C/O ratio 
of MS, S and C (N-type) stars. Also, we can explain the
M$\rightarrow$S$\rightarrow$C evolutionary sequence 
on the basis of the lighter$-$heavier s-element abundances relationship 
and the heavy-element abundances$-$C/O ratio relationship simultaneously.

\subsection * {4.2 The barium stars}

Our model 
assumes that the binary system remains always
detached, so that the only interaction between the two components of 
the binary system occurs via the accretion by the less evolved 
component of some fraction of the mass lost through stellar wind 
by the carbon- and
heavy elements-rich AGB component. In other words, the internal structure of
each component remains unaffected by the presence of its companion, so that
the usual structural properties of single stars remain valid. 
Based the
above-mentioned model, the heavy-element abundances of intrinsic AGB stars
are calculated firstly, then the heavy-element overabundances of barium
stars are self-consistently calculated through the
progressive pulsed wind accretion and mixing. 

The change equations of orbital elements
are obtained by adopting the 
angular momentum conservation of the total system 
(including the two companions and the ejected matter) 
(see Sect. 3.2.1 and Appendix). 

The observed orbital periods of 14 barium stars are in the range of 
80.53 to 6489 days. 
According to the discussions of Jorissen et al. (1998) and Zhang et al. (1999),
the barium stars with orbital period $P>1500$ or $P>1600$ days 
formed through wind accretion, 
while those with $P<600$ days formed through other scenarios.
So in calculation, we adopt 600-7000 d to be the 
orbital period range of wind accretion. 

The actual mass accretion rate between the components of binary systems
can be 0.1-1 times of B-H rate (Boffin \& Za$\check{c}$s 1994), 
while SPH simulation (Theuns et al. 1996) 
indicated that the actual rate was about 10 times 
smaller than B-H rate. We adopt 0.15 times of B-H rate as the standard
case in our calculation. 

The calculated relationships between 
the heavy-element overabundances [s/Fe] and orbital period $P$ of 
barium stars are displayed in Fig. 3, which are based on the standard case 
of wind accretion.
Here, again, the various curves correspond to different times of the neutron
exposures of the intrinsic AGB stars, and '$a$' represents the same meanings
as in Fig. 1 and Fig. 2.
Every point of every curve refers to the different orbital period,
within the range 600-7000 d. Since the shorter the orbital period, 
the greater the accretion, and hence the larger the heavy-element overabundances
will be. 
For the curves in Fig. 3, the higher points correspond to
the shorter periods. The theoretical results with $0.8\leq a \leq 2.5$
can fit the observations within the error range.

\begin{figure}
\input epsf
\epsfverbosetrue
\epsfxsize 8.8cm 
\epsfbox{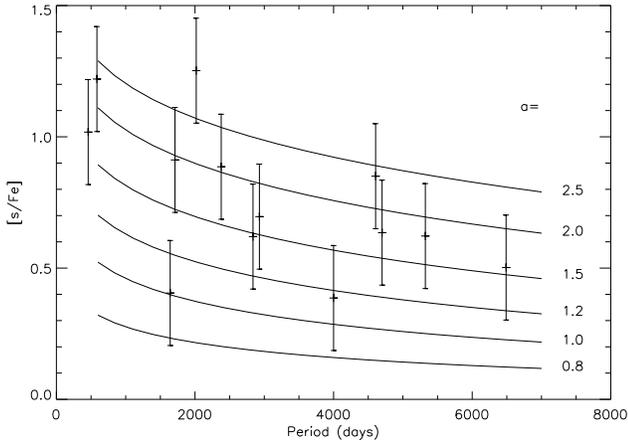} 
\caption{~Comparison of the predicted to observed relationships of 
surface heavy-element 
abundances to the orbital periods of barium stars 
corresponding to different neutron exposures of intrinsic AGB stars 
(solid lines) in standard case of wind accretion.
The numbers represent the same meanings as in Fig. 1.}
\end{figure}

To make further examination to the dependence
of heavy-element abundances on the angular momentum conservation model 
of wind accretion, the mass accretion
scenario of barium stars is analyzed carefully by comparing the predicted
heavy-element overabundances of different atomic charge Z 
with the observations 
of 14 barium stars. 

In calculation, we try to make the calculated 
eccentricity $e$ and orbital period $P$
match to the observations of the barium stars. 
The results are given in Figs.$4a-4n$.
The corresponding neutron exposure
of intrinsic AGB components, $a$, and the orbital period $P$ of
barium stars are exhibited in the every Figure. 

These results show that, for the 9 long-period barium
stars ($P$$>$1600 d), the calculated curves can fit well to the observed
heavy-element abundances in the error range (see Figs.$4a-4i$) 
according to the standard case of our wind accretion scenario.
For the two classical barium
giant stars HD 204075 ($\zeta$ Cap) and HD 16458 (HR 774), the results 
will fit the observations better after
the mass accretion rate is improved 
to 0.5 times as much as the B-H accretion rate 
(see Figs.$4j, 4k$).
For the
three barium stars with shorter orbital periods, HD 199939, HD 46407 and
HD 77247 (the orbital periods are 584.9, 457.4 and 80.53 days respectively),
the calculations 
can not fit the observations (see Figs. $4l-n$). 

Because the same program barium stars are chosen in Fig. 4 as in
Fig. 3, the reasonable result would be that the same
observational sample should correspond to the same neutron
exposure of intrinsic AGB stars in the two figures. The results
are advantageous to the standpoint: for most of barium stars, 
the corresponding neutron exposures of '$a$' in Figs.4$a-4n$ 
are consistent to the values of '$a$' 
in Fig. 3 within the error ranges. 
Namely, the comparisons between the predicted and
the observed results are basically identical in the heavy-element abundances
diagram and the heavy-element abundance$-$orbital period $P$ diagram of
barium stars.

Comparisons between the predicted and the observed
abundances of barium stars in Fig. 3 and Fig. 4 show that 
the barium stars with longer orbital
period
($P$$>$1600 d) form through wind accretion, and the change range of mass
accretion rate should be 0.1 to 0.5 times 
as much as the Bondi-Hoyle's
accretion rate; the barium stars with shorter period ($P$$<$600 d) can 
form through other scenarios: stable case C disk accretion or common envelope
ejection. Our results support quantitatively 
the conclusions of Jorissen et al. (1998) and Zhang et al. (1999). 

\section{Conclusion and discussion}

\subsection * {5.1 The intrinsic AGB stars}

Adopting the latest AGB stars evolutionary theory and nucleosynthesis scenario,
the heavy-element abundances of solar metallicity 3$M_{\odot}$ AGB stars
are calculated. 
It is shown that, adopting reasonable parameters, 
the calculated results of the heavier/lighter s-elements ratio$-$
overabundance relationships
and the C/O ratio$-$overabundance relationships can fit the observations. 
The evolution of AGB stars along the M$\rightarrow $S$\rightarrow$C
sequence is thus further explained from 
the evolutionary theories
and heavy-elements nucleosynthesis of AGB stars.
A 3$M_{\odot}$ AGB
star will undergo about 27 pulses before it becomes carbon star.

The results showed that the overabundances of s-process elements
depend significantly on the neutron exposure:
for the lower neutron exposures ($a$$<$1.5), the lighter s-process elements
are more abundant than the heavier ([hs/ls]$<$0.0); 
for the higher neutron exposures ($a$$>$1.5), 
the latter can increase strongly
with the increase of neutron exposures, and the lighter s-process elements
change slightly ([hs/ls]$>$0.0); while the results  
with $a$=1.5 can explain the solar heavy-element
abundance distribution, 
which confirm the results of Gallino et al. (1998).
Moreover, $^{12}$C abundance correlates to the s-process element
abundances. Both C/O ratio and s-element abundances increase with
the occurrence of TDU. 
After some dredge-ups, C/O ratio reaches 1,  
from which the AGB stars become carbon stars.
The observed MS, S and C (N-type)
stars lie in a range of neutron exposure: $a$=0.5-2.5. 
The over low neutron
exposure can not produce carbon stars (e.g. $a$=0.5). 
Our calculation thus provides
a further theoretical basis for the evolution of AGB stars along the
M$\rightarrow $S$\rightarrow$C sequence based on the heavy-element abundances and C/O
ratio. The general agreements between our
calculations and the observations indicate that the parameters,
the theories of evolution and nucleosynthesis scenario of AGB stars 
adopted in calculation are reasonable. 

The stars with initial main sequence mass 1.5$-$4$M_{\odot}$ can form carbon stars 
(Groenewegen et al. 1995). We consider only a 3$M_{\odot}$ star model in our
calculation.
So the comparison between the calculated results and the observations
is somewhat rough. With further studies of the evolutionary theory of AGB
stars and increasing observational data, we can expect a deeper understanding
on the nucleosynthesis of AGB stars.

\subsection * {5.2 The barium stars}

Taking the conservation of angular momentum in place of
the conservation of tangential momentum for wind accretion scenario,
considering the change of $\delta r/r$ term, 
and not neglecting the square and higher power terms of eccentricity,
the change equations of orbital
elements are recalculated.  We combine wind accretion with the nucleosynthesis
of intrinsic AGB stars, to calculate, in a self-consistent manner,
the heavy-element overabundances of barium stars through mass accretion
during successive pulsed ejection, followed by mixing.

The calculated relationships of heavy-element abundances $-$ orbital 
period $P$ can 
fit the observations within the error ranges.
Moreover, the predictions of the detailed 
abundances of different atomic charge Z can match well the observations of
11 program barium stars with longer orbital period ($P$$>$1600 d). 
The corresponding neutron exposures are in the range
of 0.8$<$$a$$<$2.5. 
The higher neutron exposure (e.g. $a$=2.0) will
produce the more abundant the heavier s-elements than the  
lighter (see Fig. 4c);
on the contrary, the abundances of the lighter s-elements are higher 
than the heavier with low neutron exposure (e.g. $a$=1.0, see Fig. 4a);
while $a$=1.5 will produce the almost equal abundances of the
heavier to the lighter s-elements (see Fig. 4h).
These neutron exposures can fit the corresponding results of 
intrinsic AGB stars.
Naturally, we can understand 
the observations of no-Tc MS and S stars, which are 
commonly believed to be the descendants of barium stars.
These results of the extrinsic AGB stars confirm the reliability
of the nucleosynthesis and evolution of intrinsic AGB stars. 
Simultaneously, the results confirm 
that our wind accretion model and parameters adopted are suitable.  

Analyzing our results, 
we understand that the barium stars with longer orbital period 
($P$$>$1600 d) form
through wind accretion.
Those with shorter orbital period ($P$$<$600 d) 
form through other scenarios,
such as dynamically stable late case C mass
transfer or common envelope ejection.
Moreover, the change
range of mass accretion rate should be 0.1 to 0.5 times as much as the
Bondi-Hoyle's accretion rate. 
The corresponding range of orbital periods and mass accretion rate 
to the formation of barium stars still need to be tested by more observations.
 
At present, the orbital elements of a large sample barium stars, 
Tc-poor S stars have been published (Udry et al. 1998a, 1998b; Carquillat et al. 1998).
But the corresponding
heavy-element abundances have not been obtained. So we need the high
resolution, high signal-to-noise spectral observations of these stars, 
which are combined
with the observations of orbital elements,
to research their characters and formation. 

In addition, we should note that
metallicity is an important factor to the AGB stars nucleosynthesis
and the formation of chemical peculiar stars.
Actually, $^{13}$C neutron source is related to the metallicity 
(Busso et al. 1995, 1999 and references therein; Gallino et al. 1999).
Gallino et al. (1998) calculated the s-element nucleosynthesis of 
2$M_{\odot}$ AGB stars with low metallicity $Z$=0.01, and 
obtained similar abundance distribution to the 
3$M_{\odot}$ with solar metallicity AGB stars model. 
Busso et al. (1999) and Zhang et al. (1998b) 
have discussed the inverse correlation
between the heavy-element abundances and the metallicity [Fe/H] 
of intrinsic and extrinsic AGB stars. 
Also, the nucleosynthesis results of low metallicity AGB stars
are more suitable to study the Galactic chemical evolution 
in the early stage of the Galaxy.
For extrinsic AGB stars,
Jorissen et al. (1998) suggested that 
the s-process operation was more efficient in a low-metallicity population, so
the Pop. II CH stars may has accreted the material much enriched in heavy 
elements from the former AGB companion.
In this paper, our main aims are 
(1) calculating the AGB stars nucleosynthesis, so that we can explain the 
observed heavy-element abundances of MS, S and C (N-type) stars, 
which are near solar metallicity, and supply evidence
to the M$\rightarrow $S$\rightarrow$C evolutionary sequence; 
(2) discussing the wind accretion scenario of Ba stars, 
which are near solar metallicity too (Za$\check{c}$s 1994; 
Smith et al. 1993). 
So we 
only calculate the solar metallicity case. 
We will extend to study the
low metallicity case in the forthcoming paper.

\begin{acknowledgements}
We thank Dr. Oscar Straniero for very useful suggestions 
on the original manuscript, and on the finally improved presentation.
We thank Dr. Maurizio Busso for fruitful discussion 
and sending important material to us. 
We thank Dr. Peng Qiuhe and Dr. Ma Jun for the interesting 
discussions.
Thank Dr. Jiang Biwei, Liu Junhong, Li Ji, Zhang Yanxia, Shi Jianrong 
and Luo Ali for their friendly help.
This research work is supported by the National 
Natural Science Foundation of
  China under grant No. 19725312, No. 19973002, and the 
  Major State Basic Research Development Program. 
\end{acknowledgements}


{\bf Appendix}

{\bf The angular momentum conservation model of wind accretion  for barium
stars: Equations}  

For the binary system, the two components
(an intrinsic AGB star with mass $M_{1}$, the present white dwarf, 
and a main sequence star with mass $M_{2}$,
the present barium star)
 rotating around the mass core C, so the total angular momentum
is conservative in the mass core reference frame. 
If the two components exchange material through wind
accretion, the angular momentum conservation of total system is showed by:
\begin{equation}
\Delta(\mu r^{2}\dot{\theta})=\omega r_{1}^{2}(\Delta M_{1}+\Delta M_{2})
+r_{2}v(\Delta M_{1}+\Delta M_{2}),
\end{equation}
where $\mu$ is
reduced mass, and $r$ is the distance from $M_{2}$ to $M_{1}$.
$r_{1}$ and $r_{2}$ are the distances from $M_{1}$, $M_{2}$ to the
mass core C respectively.
$\omega$ (={2$\pi /P$) is angular velocity, where 
$P=2 \pi A^2(1-e^2)^{1 \over 2}/h$ 
is the orbital period (Huang 1956).  
$v$ is an additional effective velocity defined through
the angular momentum variation in the direction of orbital motion of component 2.
The first term on the right side of the equal-sign is
the angular momentum lost by the escaping material and the second term
is the additional angular momentum lost by the escaping material.

For the binary system, according to Huang (1956), the changes of
orbital elements, the orbital semi-major axis $A$ and eccentricity $e$, are

\begin{equation}
\frac {\Delta A}{A}=\frac {\Delta(M_1+M_2)}{M_1+M_2}-\frac {\Delta E}{E},
\end{equation}
\begin{equation}
\frac {e\Delta e}{1-e^2}=\frac {\Delta(M_1+M_2)}{M_1+M_2}-
{1\over2} \frac {\Delta E}{E}-\frac {\Delta h}{h},
\end{equation}

where
\begin{eqnarray}
\frac {\Delta E}{E} & = & \frac {\Delta T}{E}+\frac {\Delta \Omega}{E}  \nonumber\\
    & = & \frac {r\dot {\theta}}{E} \Delta (r\dot {\theta})
          + \left [\frac {\Delta (M_{1}+M_{2})}{M_{1}+M_{2}}\frac{2A}{r}
          -\frac {2A\Delta r}{r^2} \right ]~.  
\end{eqnarray}

According to the angular momentum conservation model, 
the $\Delta (r\dot {\theta})$ term can be obtained from the following 
equation:
\begin{eqnarray}
\frac{M_1M_2}{M_1+M_2}r\Delta (r\dot {\theta}) 
            & = & \Delta \left ({\frac{M_1M_2}{M_1+M_2}r~r\dot {\theta}} \right ) \nonumber\\
            & - & \Delta \left ({\frac{M_1M_2}{M_1+M_2}r} \right )~r\dot {\theta}  \nonumber\\
            & = & \Delta(\mu r^{2}\dot{\theta})
                  -\Delta \left ({\frac{M_1M_2}{M_1+M_2}r} \right )~r\dot {\theta}    \nonumber\\
            & = &  \omega r_{1}^{2} (\Delta M_{1}+\Delta M_{2})            \nonumber\\
            & + & r_{2}v(\Delta M_{1}+\Delta M_{2})                      \nonumber\\
            & - & \Delta \left ({\frac{M_1M_2}{M_1+M_2}r} \right )~r\dot {\theta}, 
\end{eqnarray}

thus
\begin{eqnarray}
\Delta (r\dot {\theta}) & = & -r\dot {\theta} \left [\frac {\Delta {M_1}}{M_1}
                          +\frac {\Delta {M_2}}{M_2}
                          -\frac {\Delta (M_1+M_2)}{M_1+M_2} \right ]
                          -r\dot {\theta} \frac {\Delta r}{r} \nonumber\\
                    & + & \frac {\Delta (M_{1}+M_{2})}{M_{2}}v   
                         +\frac {rG^ {1\over2}M_{2}\Delta (M_{1}+M_{2})}
                          {M_{1}(M_{1}+M_{2})^ {1\over2}A^{3\over2}}~,
\end{eqnarray}

then, we can obtain
\begin{eqnarray}
\frac {r\dot {\theta}\Delta (r\dot {\theta})}{E} 
           & = & -\frac {(r\dot {\theta})^2}{E} \left [\frac {\Delta {M_1}}{M_1}
                          +\frac {\Delta {M_2}}{M_2}
                          -\frac {\Delta (M_1+M_2)}{M_1+M_2} \right ] \nonumber\\
                  & - & \frac {r\dot {\theta}^2}{E} \Delta r 
                          +\frac {r\dot {\theta}}{E}\frac {\Delta (M_{1}+M_{2})}{M_{2}}v   \nonumber\\
                  & + &  \frac {r^2\dot {\theta}}{E}\frac {G^ {1\over2}M_{2}\Delta (M_{1}+M_{2})}
                          {M_{1}(M_{1}+M_{2})^ {1\over2}A^ {3\over2}}    \nonumber\\
           & = & 2(1-e^2)^ {1\over2}
                      \left [\frac {\Delta {M_1}}{M_1}+\frac {\Delta {M_2}}{M_2}  \right ]   \nonumber\\                          
                  & - & 2(1-e^2)^ {1\over2}\frac {\Delta (M_{1}+M_{2})}{M_2} \frac {v} {v_{\rm orb}} \nonumber\\
                  & - & 2(1-e^2)^ {1\over2}\frac {M_2 \Delta (M_1+ M_2)}{M_1(M_1+M_2)}  \nonumber\\
                  & - & 2(1-e^2)^ {1\over2}\frac {\Delta(M_1+M_2)}{M_1+M_2}        
                      +\frac {2\Delta r}{A(1-e^2)^{1\over2}}~.
\end{eqnarray}

Thus
\begin{eqnarray}
\frac {\Delta E}{E}  & = & 2(1-e^2)^ {1\over2}         
                      \left [\frac {\Delta M_1}{M_1}+\frac {\Delta M_2}{M_2}
                       -\frac {\Delta (M_{1}+M_{2})}{M_2} \frac {v} {v_{\rm orb}} \right ] \nonumber\\ 
                  & - & 2(1-e^2)^ {1\over2} \frac {M_2 \Delta (M_1+ M_2)}{M_1(M_1+M_2)}    \nonumber\\                  
                  & + & (2-2(1-e^2)^ {1\over2})\frac {\Delta(M_1+M_2)}{M_1+M_2} ~.        
\end{eqnarray}

The $\Delta h/h$ term is:
\begin{eqnarray}
\frac {\Delta h}{h} & = & -\frac {\Delta M_{1}}{M_{1}}-\frac {\Delta M_{2}}{M_{2}}                                                  
                     +  \frac {2+e^2}{2(1-e^2)^{1\over2}} \frac {\Delta (M_1+ M_2)}{M_{2}}
                     {v\over v_{\rm orb}}                       \nonumber\\  
          & + & \frac {M_2 \Delta (M_{1}+M_{2})}{M_{1}(M_{1}+M_{2})} 
           +\frac {\Delta (M_{1}+M_{2})}{M_{1}+M_{2}}  ~. 
\end{eqnarray}

Combining Eqs. (9), (10), (15) and (17), we can obtain the changes of orbital 
semi-major axis $A$ and eccentricity $e$:

\begin{eqnarray}
\frac{\Delta A}{A} & = &  - 2(1-e^2)^{1\over2} 
                 \left [\frac{\Delta M_1}{M_1}+\frac{\Delta M_2}{M_2}
                 -  \frac {\Delta M_1+ \Delta M_2}{M_2} \frac {v} {v_{\rm orb}} \right ] \nonumber \\
                & + & 2(1-e^2)^{1\over2}\frac {M_2 (\Delta M_1+ \Delta M_2)}{M_1(M_1+M_2)}  \nonumber \\
                & + & [2(1-e^2)^{1\over2}-1] \frac {\Delta M_1+ \Delta M_2}{M_1+M_2},            
\end{eqnarray}

\begin{eqnarray}
\frac{e\Delta e}{1-e^2} & = &  [1-(1-e^2)^ {1\over2}]
               \left [\frac {\Delta M_1}{M_1}+\frac {\Delta M_2}{M_2}         
                - \frac {\Delta M_1+ \Delta M_2}{M_1+M_2} \right ] \nonumber \\
              & - & [1-(1-e^2)^ {1\over2}] \frac {M_2 (\Delta M_1+ \Delta M_2)}{M_1(M_1+M_2)}  \nonumber \\
              & - & \frac {3e^2}{2(1-e^2)^{1\over2}} \frac {\Delta
              M_1+ \Delta M_2}{M_2} \frac {v}{v_{\rm orb}}.
\end{eqnarray}

\newpage

\begin{figure}
\input epsf
\epsfverbosetrue

\epsfxsize 8.8cm  
\epsfbox{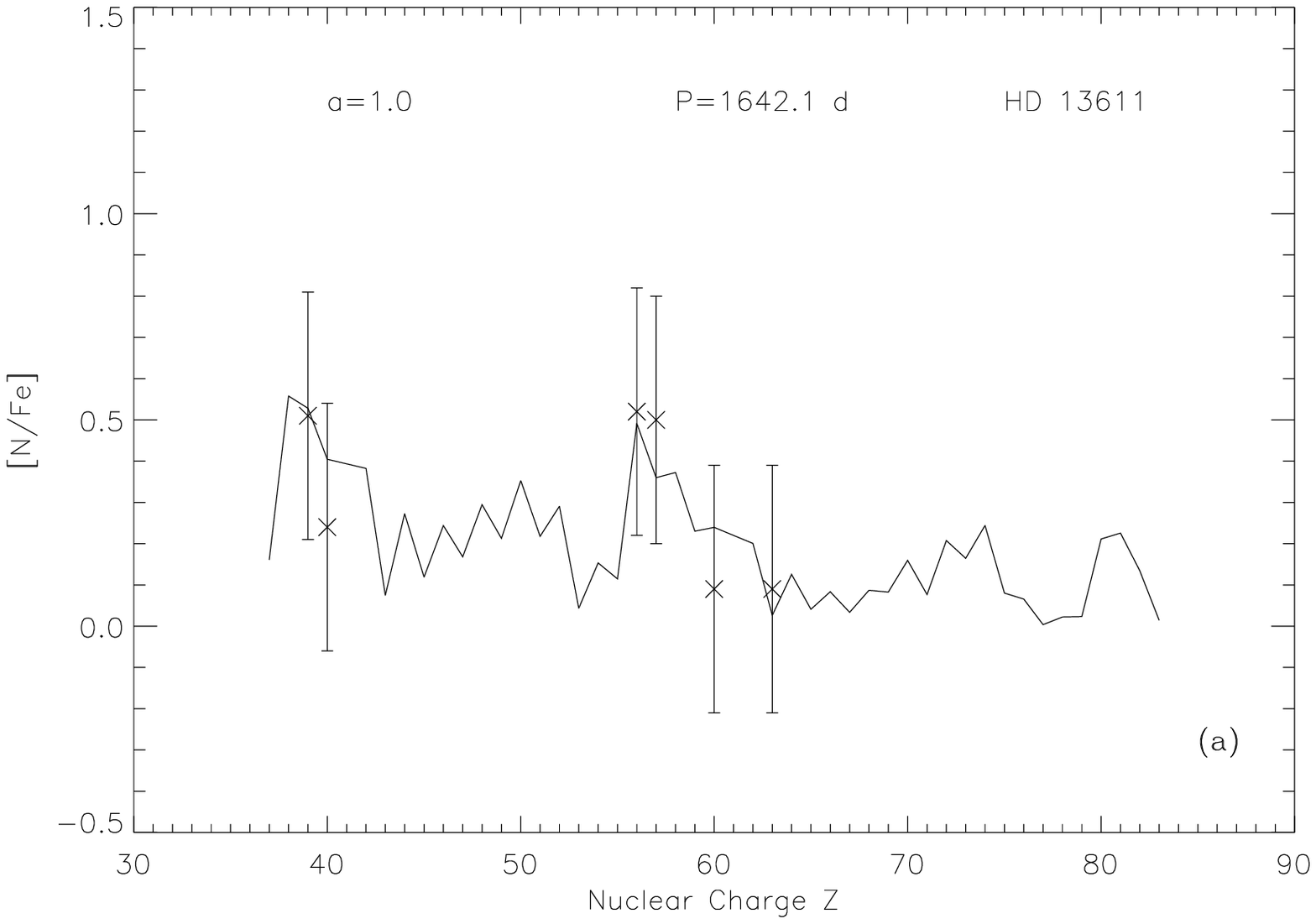}

\vspace*{10mm}

\epsfxsize 8.8cm 
\epsfbox{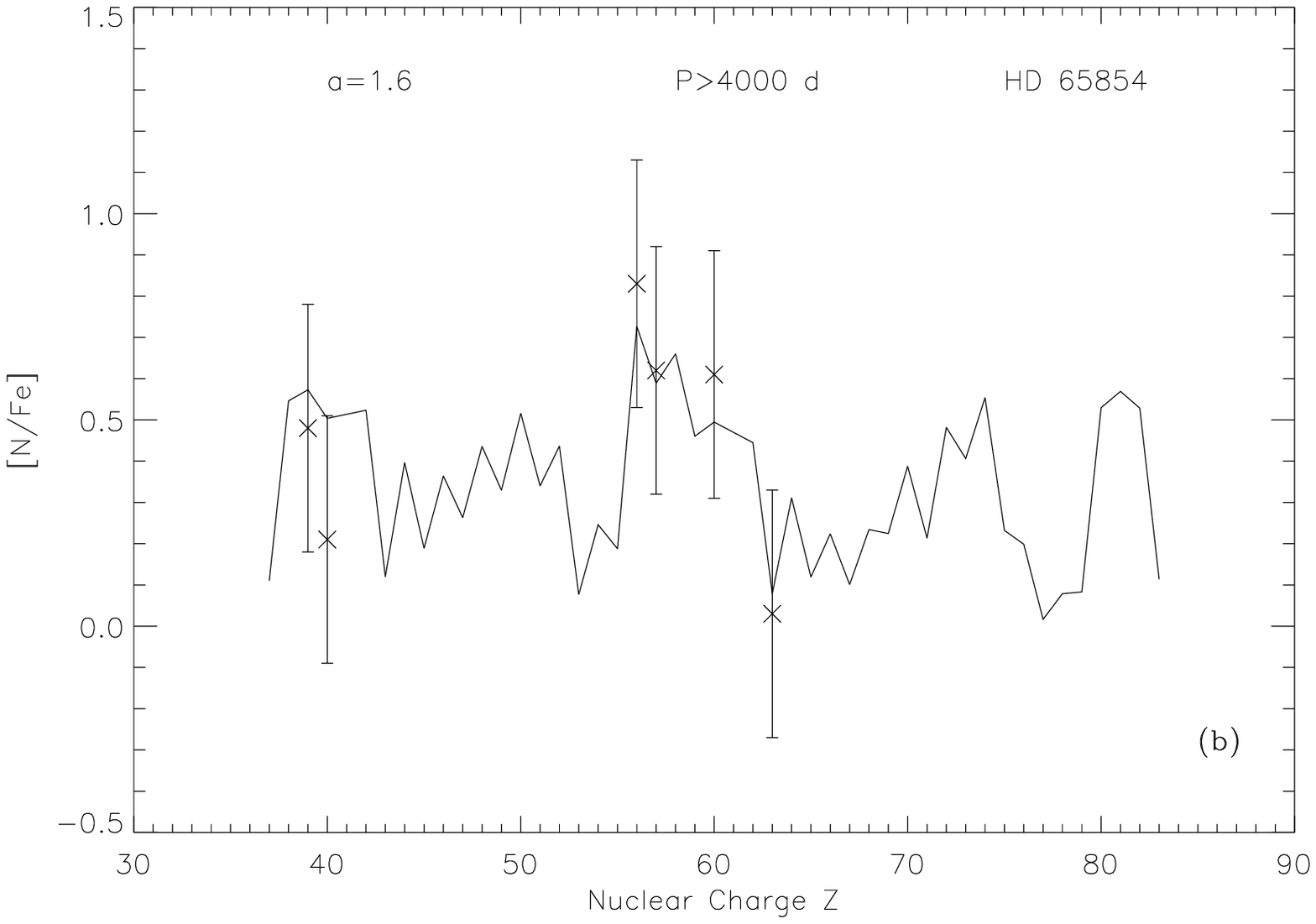} 

\vspace*{10mm}

\epsfxsize 8.8cm  
\epsfbox{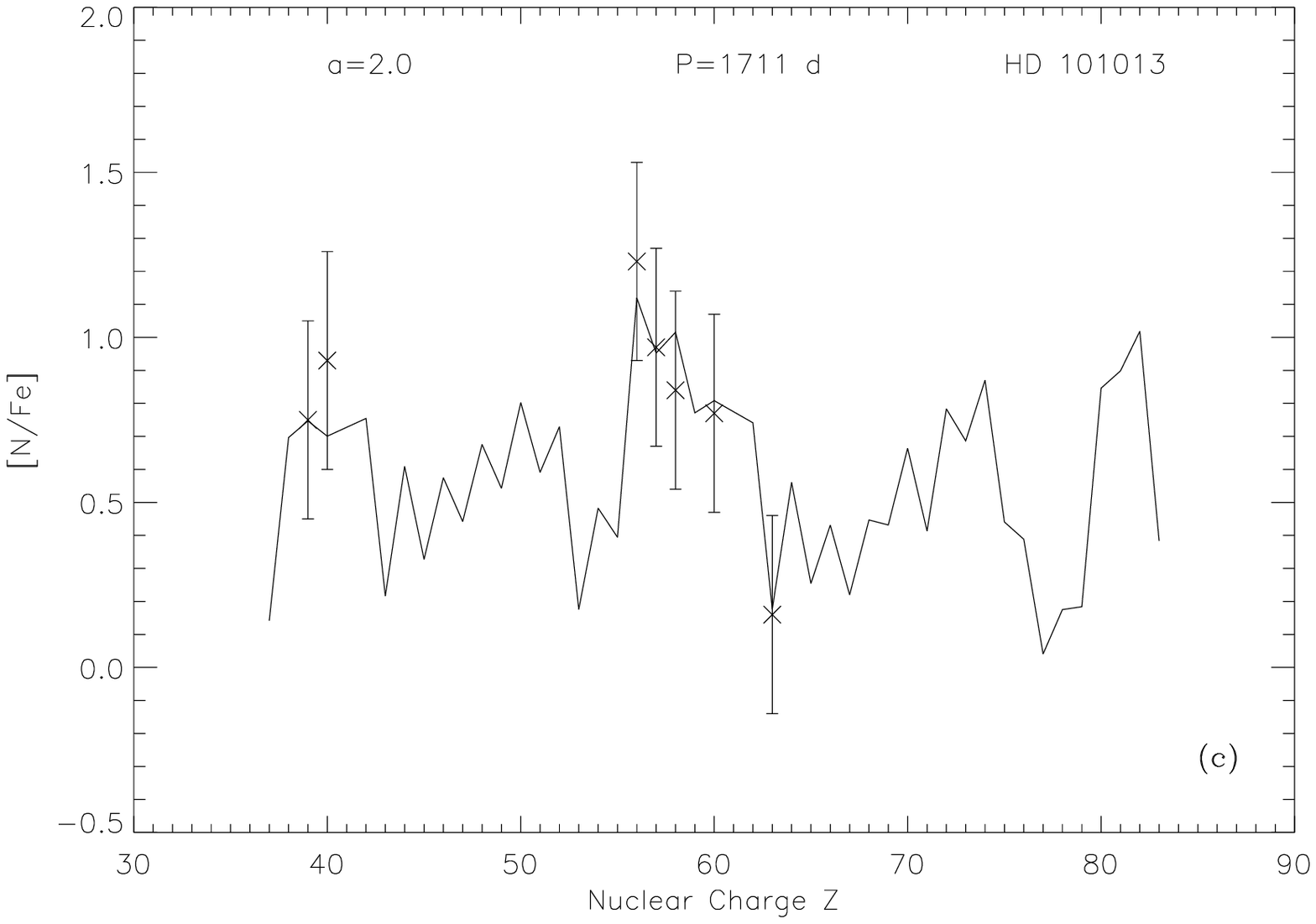}

\end{figure}

\begin{figure}
\input epsf
\epsfverbosetrue

\epsfxsize 8.8cm 
\epsfbox{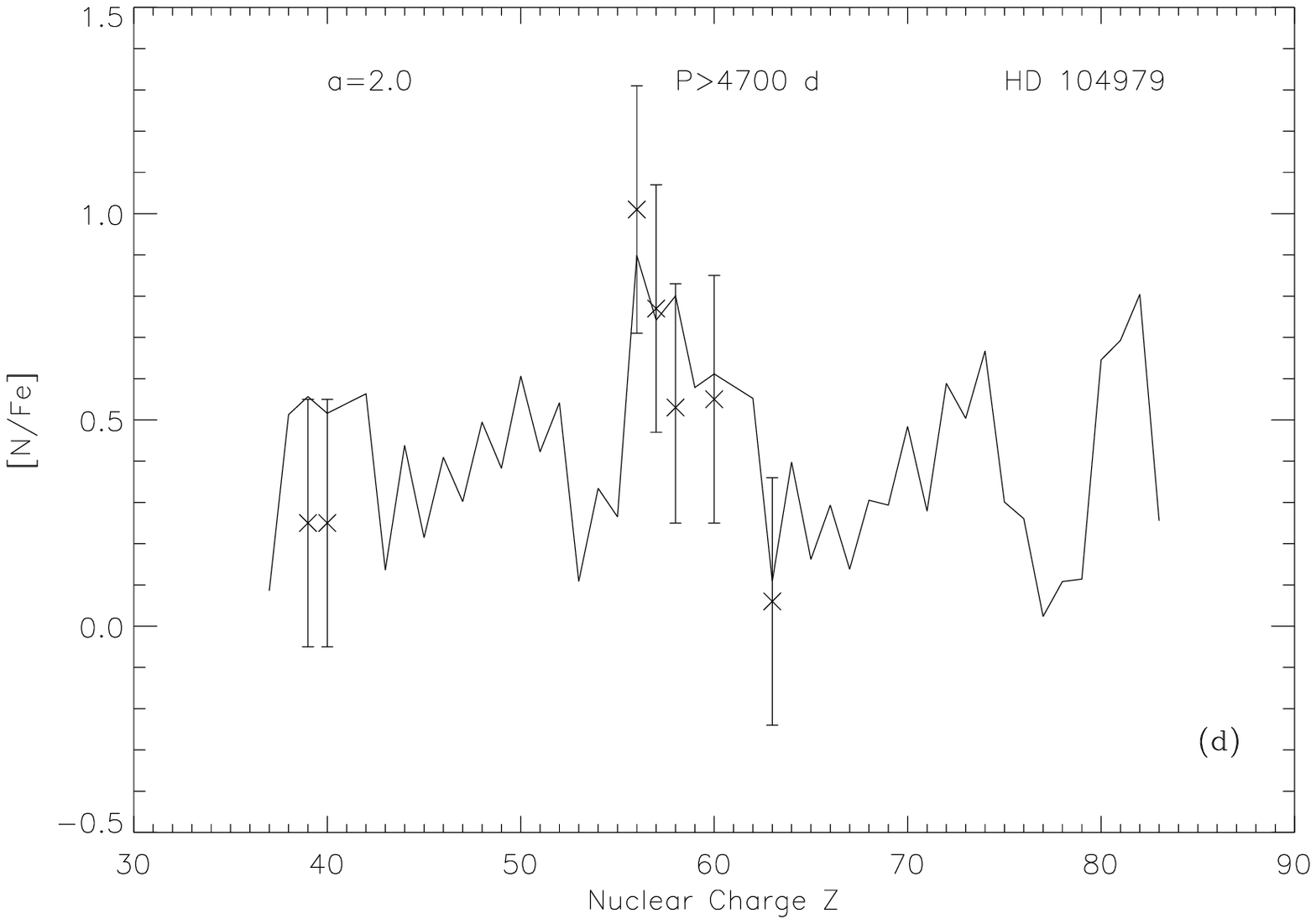} 

\vspace*{10mm}

\epsfxsize 8.8cm  
\epsfbox{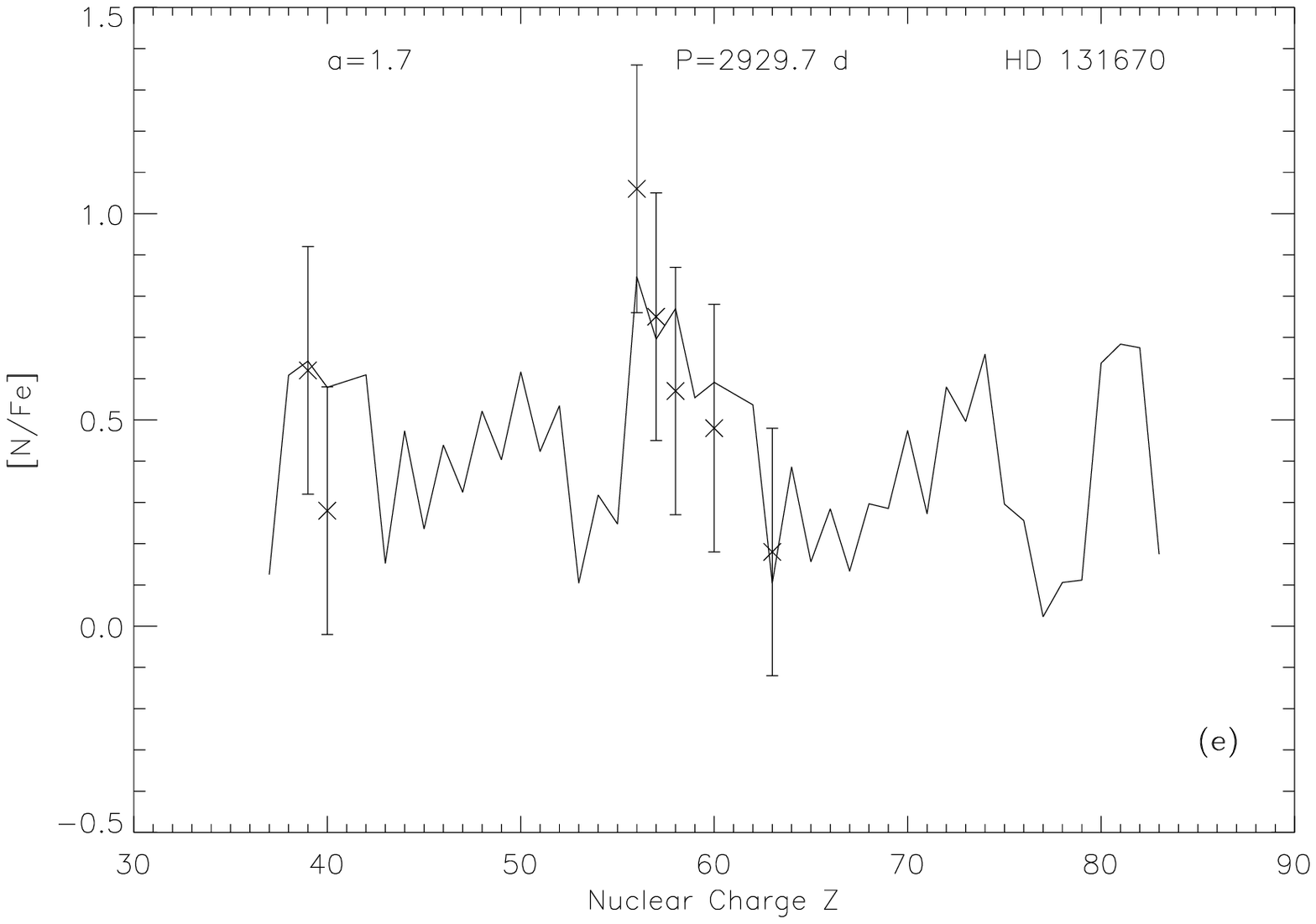} 

\vspace*{10mm}

\epsfxsize 8.8cm  
\epsfbox{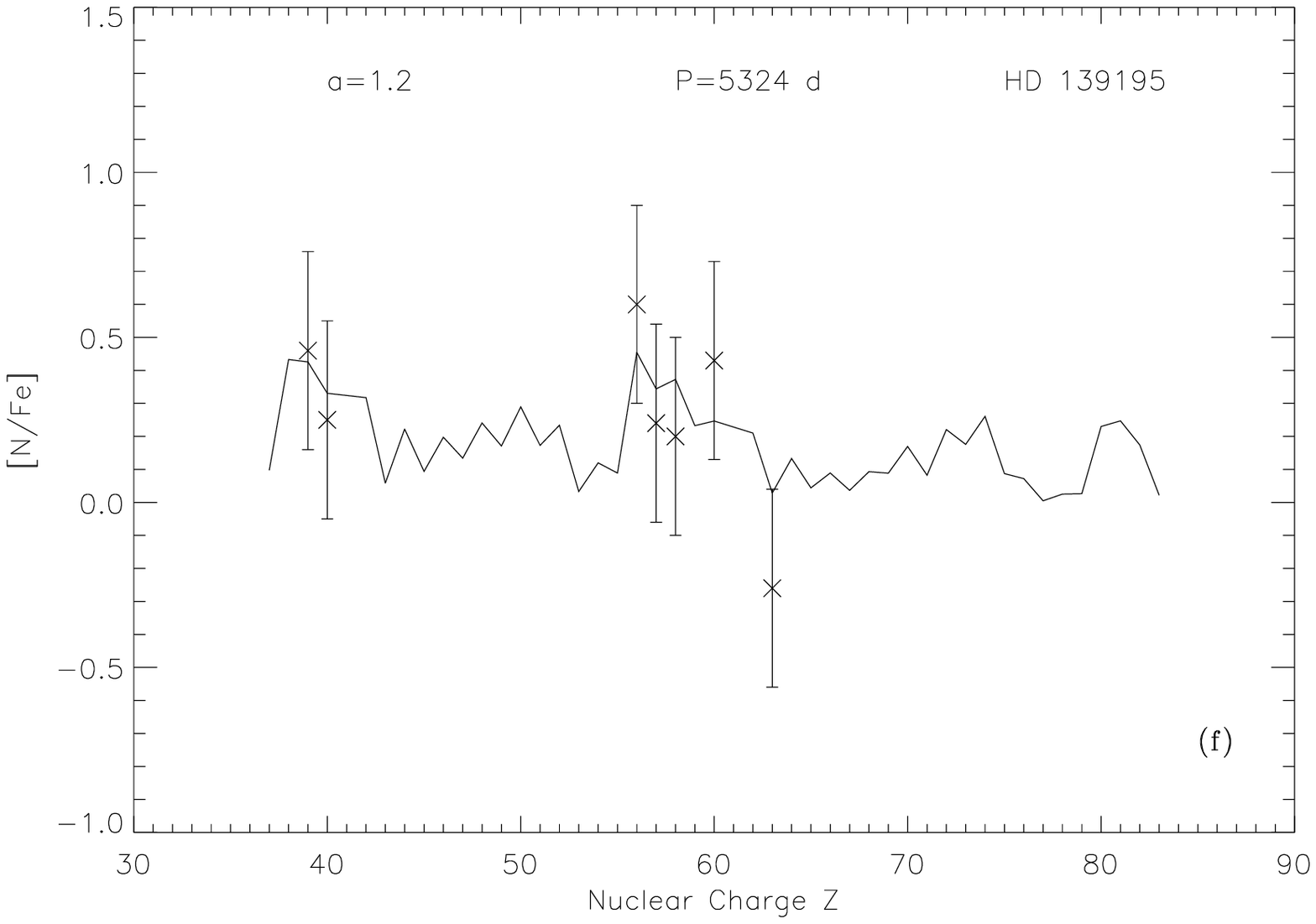} 
 
\end{figure}

\begin{figure}
\input epsf
\epsfverbosetrue

\epsfxsize 8.8cm 
\epsfbox{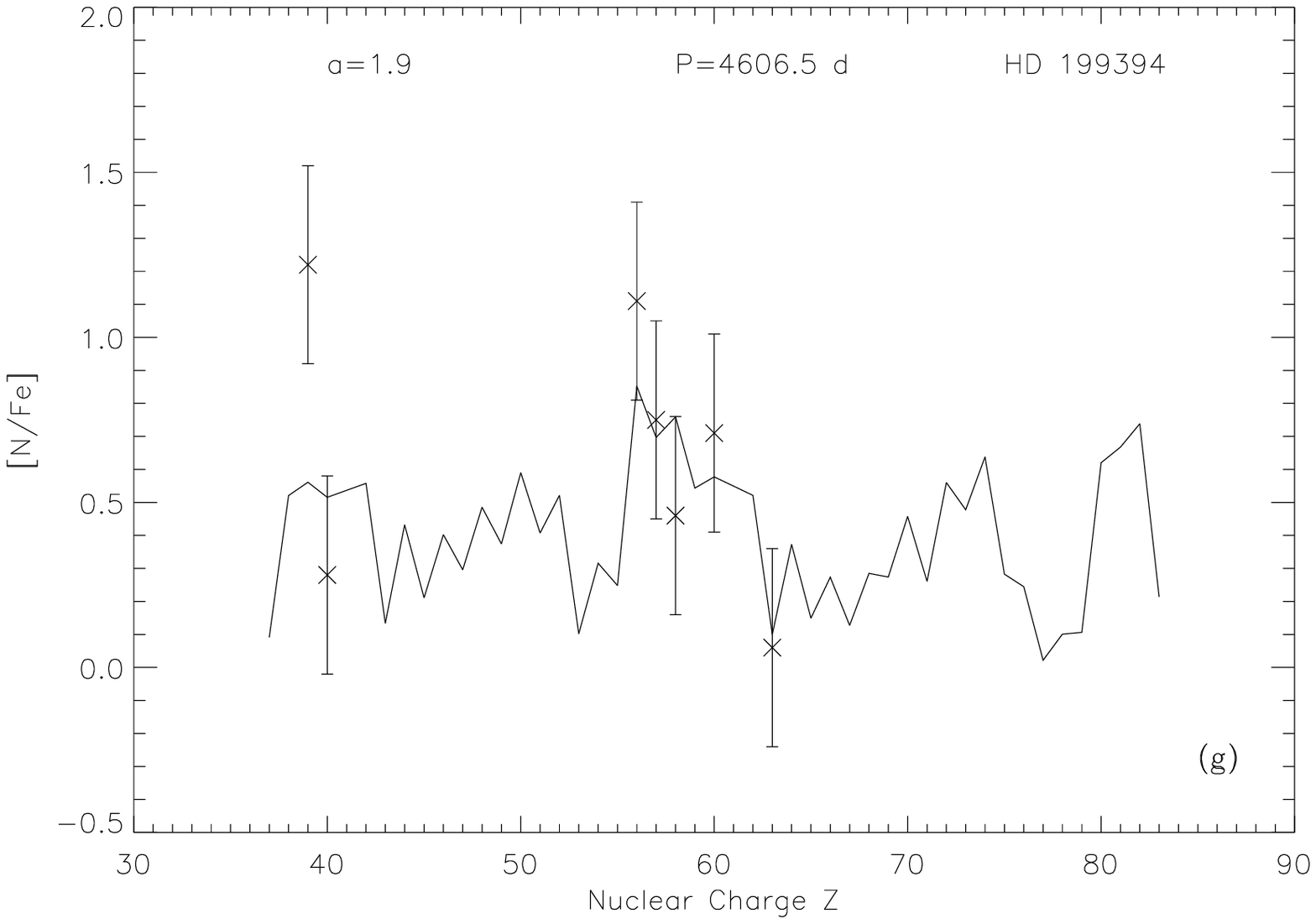} 

\vspace*{10mm}

\epsfxsize 8.8cm  
\epsfbox{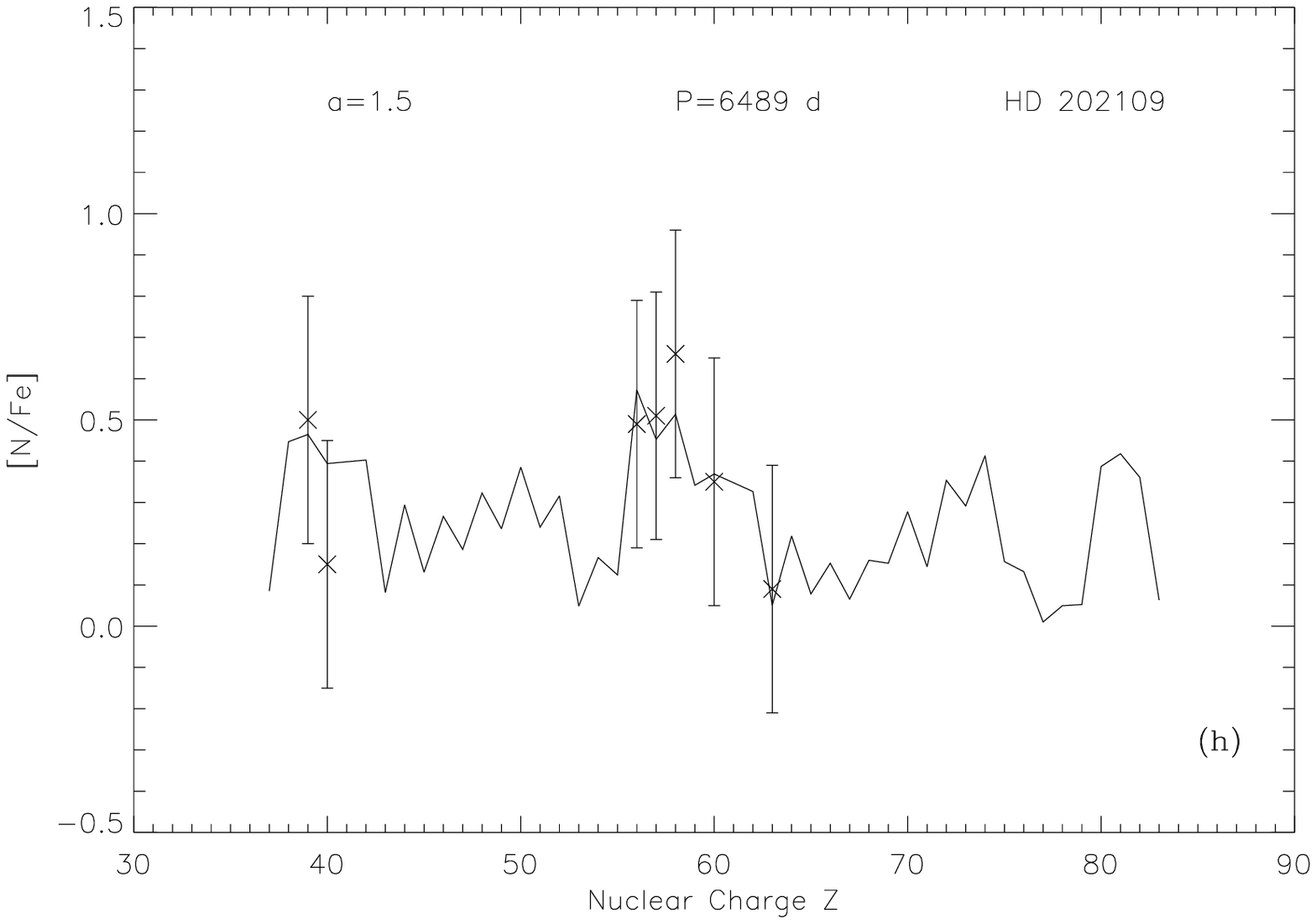} 

\vspace*{10mm}

\epsfxsize 8.8cm  
\epsfbox{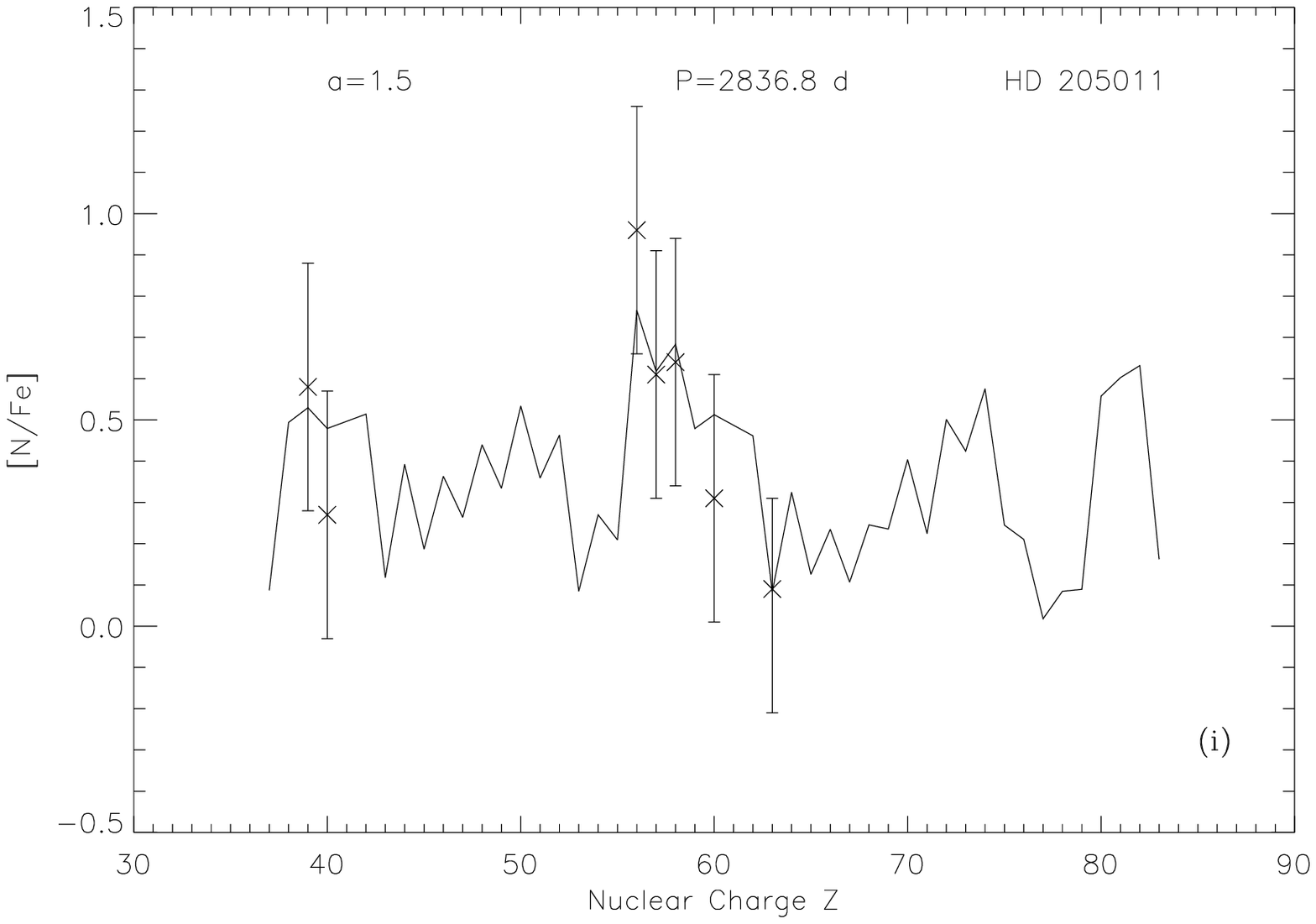} 
 
\end{figure}

\begin{figure}
\input epsf
\epsfverbosetrue

\epsfxsize 8.8cm  
\epsfbox{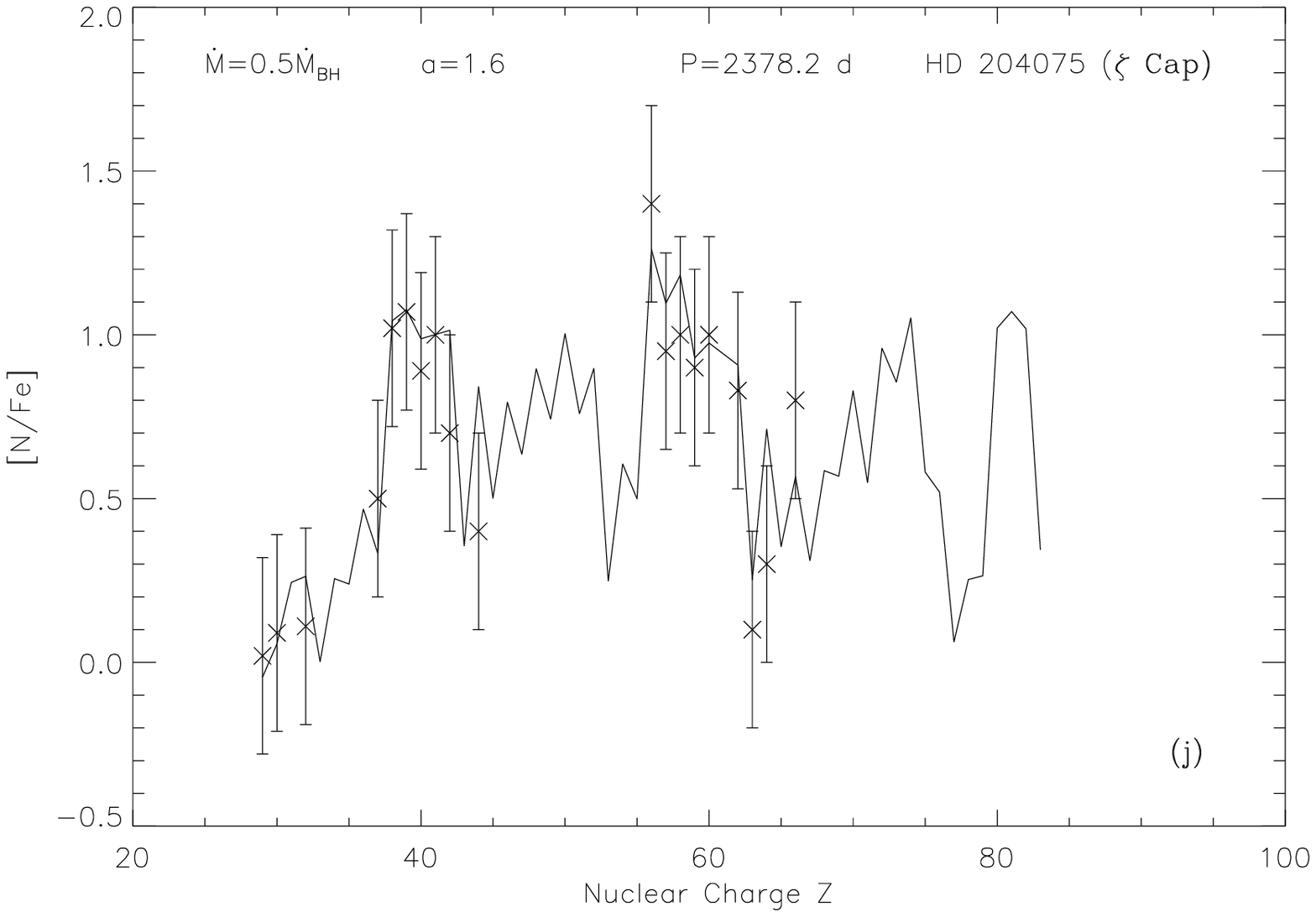}

\vspace*{10mm}

\epsfxsize 8.8cm 
\epsfbox{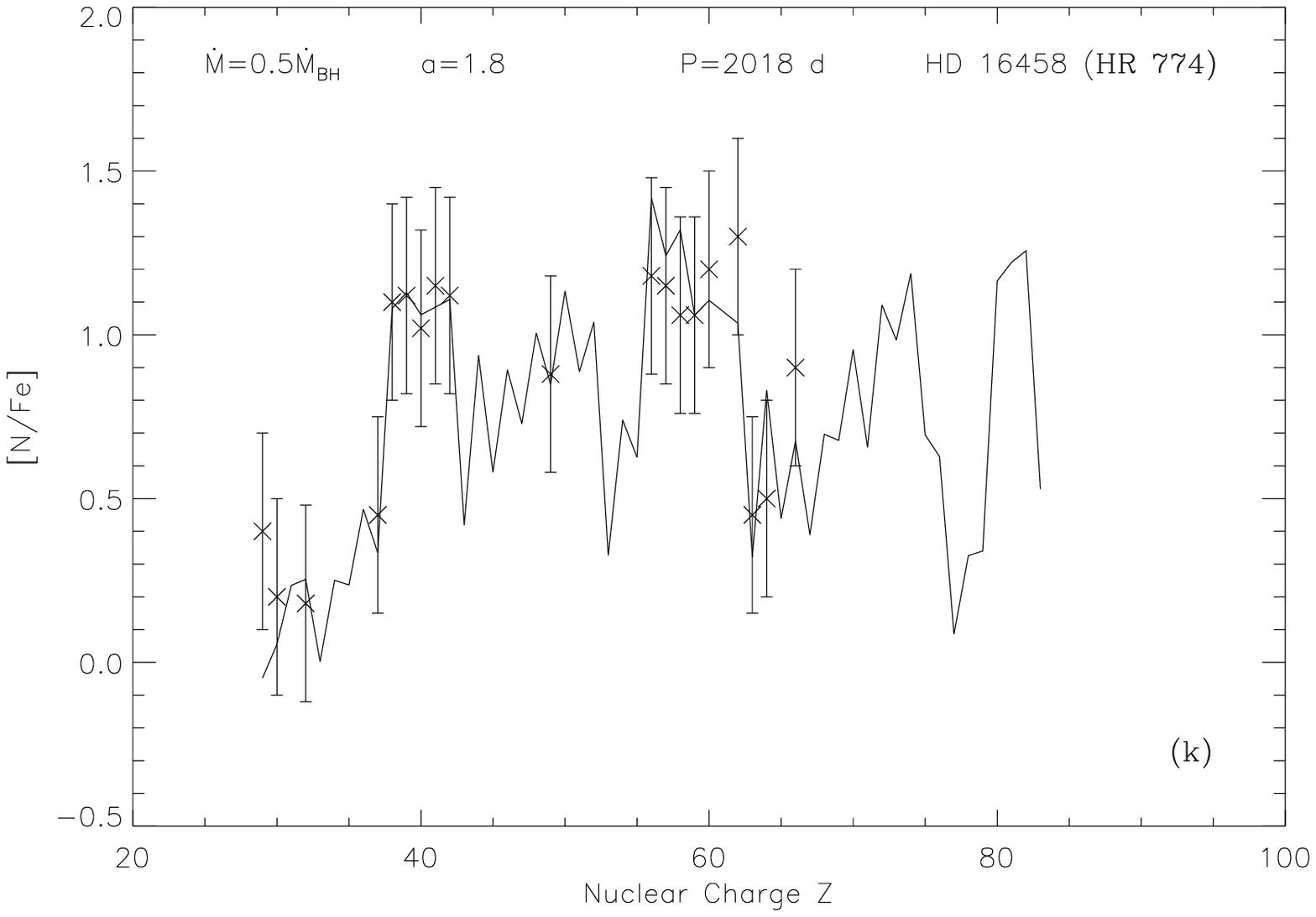} 

\end{figure}


\begin{figure}
\input epsf
\epsfverbosetrue

\epsfxsize 8.8cm 
\epsfbox{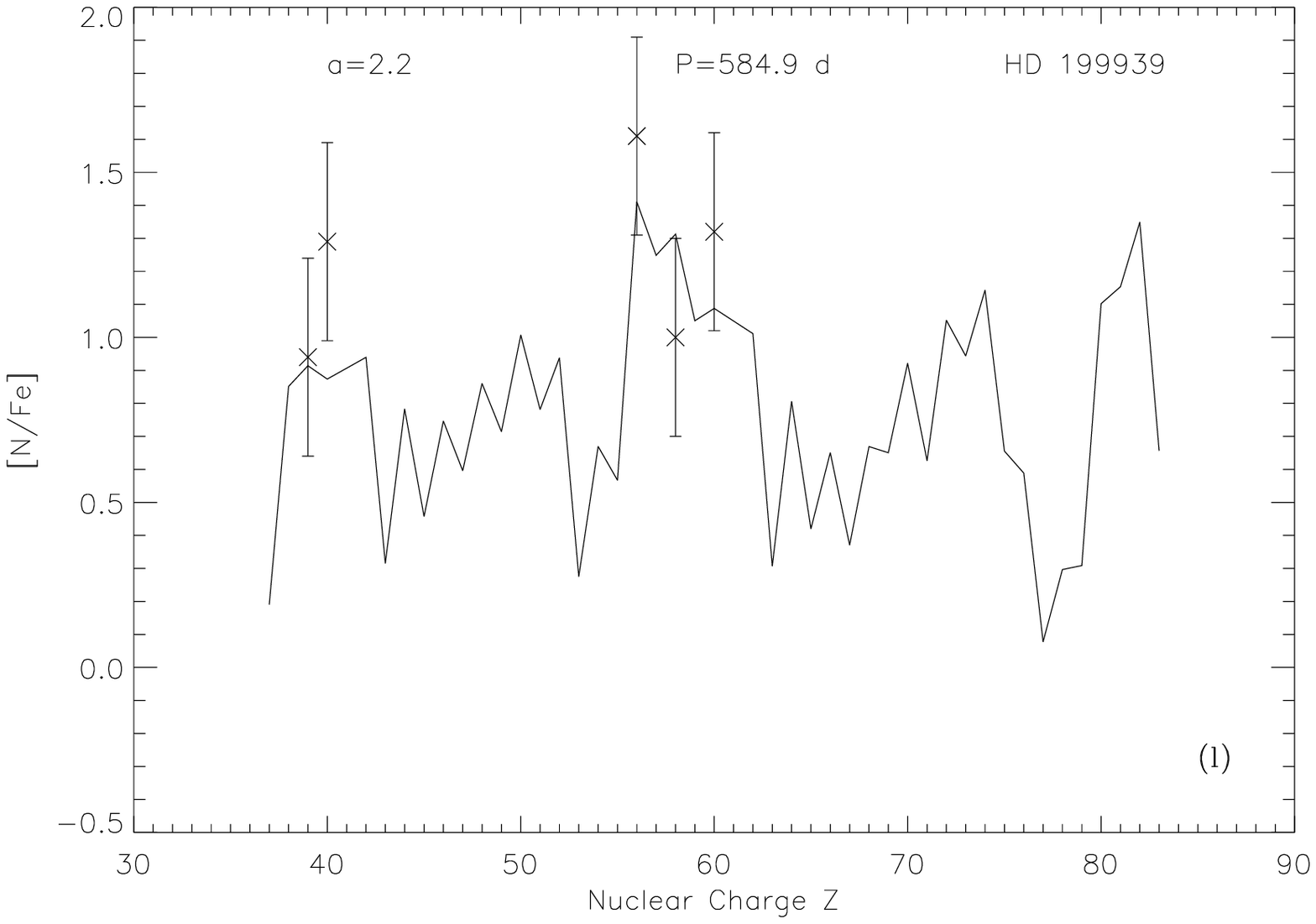}

\vspace*{10mm}

\epsfxsize 8.8cm 
\epsfbox{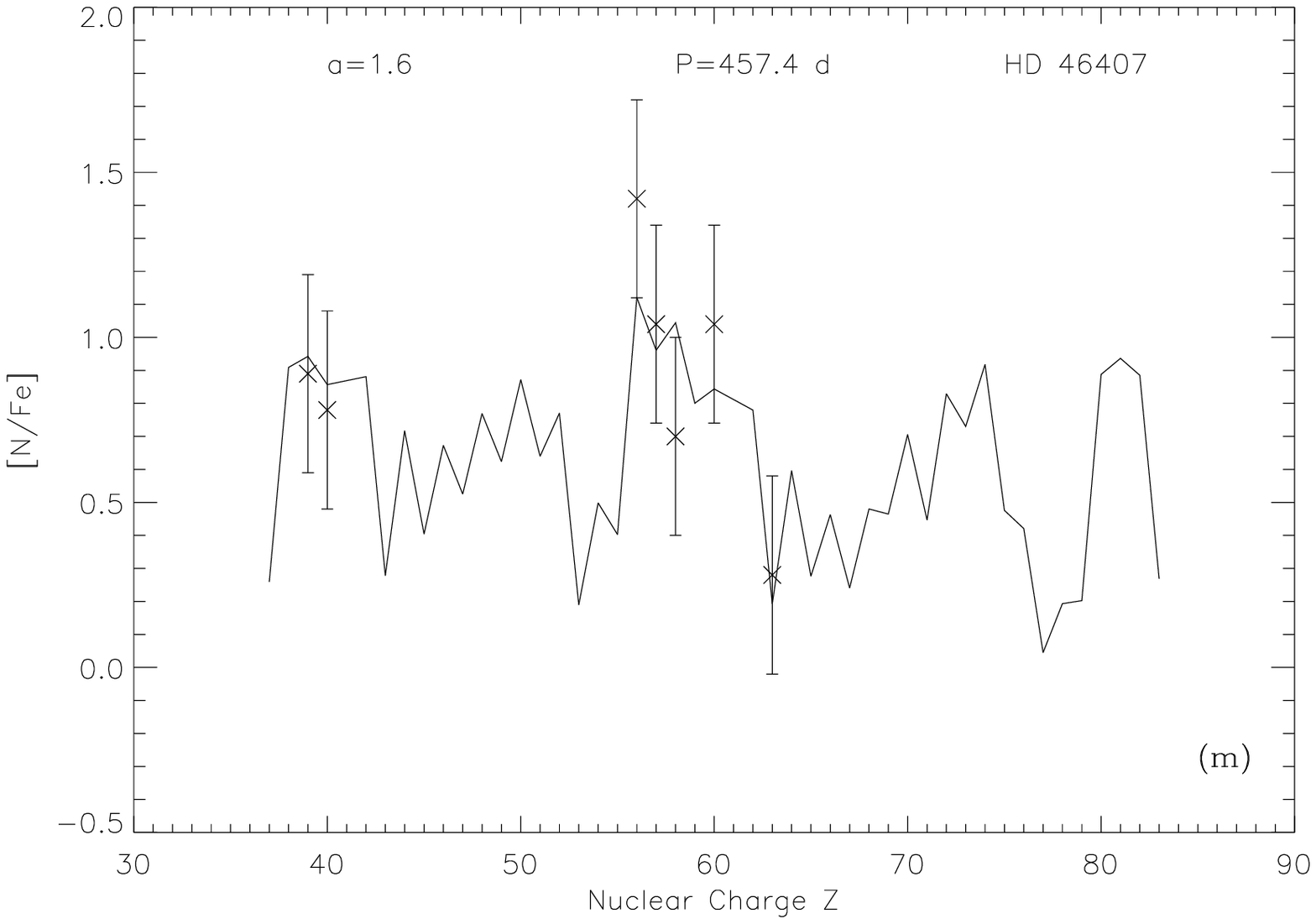} 

\vspace*{10mm}

\epsfxsize 8.8cm 
\epsfbox{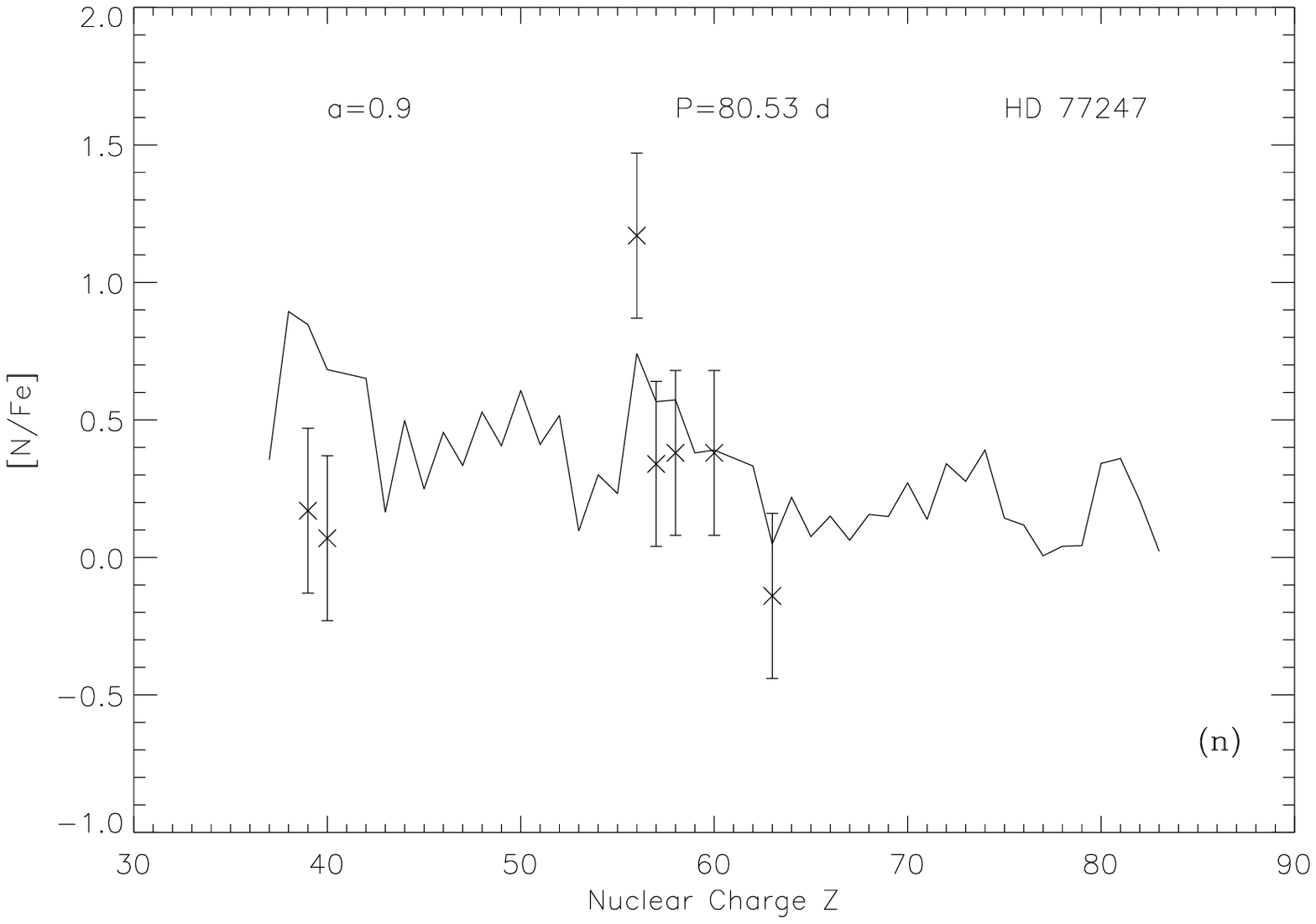} 

\caption{The fitting of the predicted to observed heavy-element 
abundances of 14 barium stars in standard case of wind accretion. 
But the curves in Fig. 4j and k represent the results of higher accretion
rate: 0.5 times of the Bondi-Hoyle's rate. 
In every figure, the alphabet '$a$' represents the times of the 
corresponding neutron exposure in the $^{13}$C profile suggested by the Fig. 1 of 
Gallino et al. (1998), and 'P' represents the orbital period of the 
barium star.}

\end{figure}

\end{document}